\documentclass[apj]{emulateapj}
\usepackage{apjfonts, hyperref}
\hypersetup{draft=true}
\slugcomment{Accepted for publication to ApJ}
\shorttitle{The Milky Way's Kiloparsec Scale Wind}
\shortauthors{Everett et al.}

\newcommand{\alfven}{{Alfv\'en}~}

\newcommand{\al}[2]{#1\,\textsc{#2}}

\begin{document}

\bibliographystyle{apj}

\title{The Milky Way's Kiloparsec Scale Wind: \\ A Hybrid Cosmic-Ray and
  Thermally Driven Outflow}

\author{by John E. Everett\altaffilmark{1,2,3}, 
  Ellen G. Zweibel\altaffilmark{1,2,3}, 
  Robert A. Benjamin\altaffilmark{4}, \\
  Dan McCammon\altaffilmark{2}, 
  Lindsay Rocks\altaffilmark{2},\\
  John S. Gallagher, III\altaffilmark{1}}
\altaffiltext{1}{University of Wisconsin--Madison, Department of
  Astronomy, 475 N. Charter, Madison, WI 53706}
\altaffiltext{2}{University of Wisconsin--Madison, Department of
  Physics, 425 N. Charter, Madison, WI 53706} 
\altaffiltext{3}{Center for Magnetic Self-Organization in Laboratory and Astrophysical Plasmas}
\altaffiltext{4}{University of Wisconsin--Whitewater, Department of
  Physics, 800 West Main Street, Whitewater, WI 53190}
\email{everett@physics.wisc.edu}

\begin{abstract}
We apply a wind model, driven by combined cosmic-ray and thermal-gas
pressure, to the Milky Way, and show that the observed Galactic
diffuse soft X-ray emission can be better explained by a wind than by
previous static gas models.  We find that cosmic-ray pressure is
essential to driving the observed wind.  Having thus defined a
``best-fit'' model for a Galactic wind, we explore variations in the
base parameters and show how the wind's properties vary with changes
in gas pressure, cosmic-ray pressure and density.  We demonstrate the
importance of cosmic rays in launching winds, and the effect cosmic
rays have on wind dynamics.  In addition, this model adds support to
the hypothesis of Breitschwerdt and collaborators that such a wind may
help explain the relatively small gradient observed in $\gamma$-ray
emission as a function of galactocentric radius.
\end{abstract}


\section{Introduction}\label{Intro}

Large-scale galactic outflows are usually considered in the context of
starburst galaxies or Active Galactic Nuclei \citep{VeCeBH2005}.
These outflows are interesting not only intrinsically (what drives the
outflow?) but for the interstellar and intergalactic media (how is the
host galaxy affected, and what metals are ejected from the galaxy?).

To examine these questions, we have built a thermal and cosmic-ray
driven wind model.  Our investigation into such models was first
inspired by observational hints that the Milky Way may possess a
kiloparsec-scale wind; this paper further explores that possibility.
To motivate this study, we first introduce the observational evidence
for a Galactic wind (\S\S\ref{diffuseXrays} and \ref{CRdensity}) and
then introduce the cosmic-ray and thermally driven wind model
(\S\ref{modelDef}).  In \S\ref{compareToObs}, we calculate the X-ray
emission from this wind, and then compare it to the observations.
After finding the best-fit wind model, we then explore the parameter
space around that model (\S\ref{ParameterSurvey}) to understand more
about how the wind is modified by varying the input and fit
parameters.  Our conclusions are given in \S\ref{Conclusions}.

But first, observational hints for an outflow from our own Galaxy.

%To start applying this model, we look to a perhaps
%unlikely first suspect: the Milky Way.

\subsection{X-ray Observations}\label{diffuseXrays}

\begin{figure*}[]
\begin{center}
\includegraphics[width=14cm]{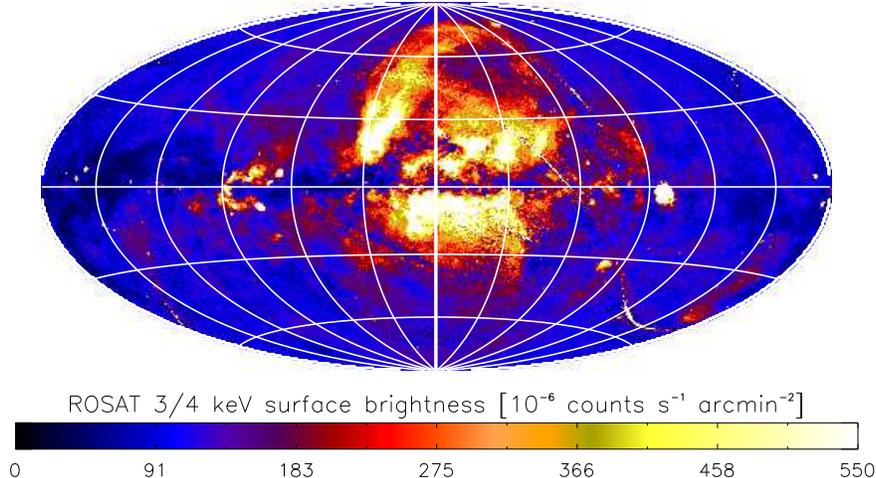}
\caption{X-ray emission at $3/4$~keV (the ``R45 band'') as seen by
  \textit{ROSAT} \citep{SnowdenEtAl97}.  These observations suggest a
  ``Galactic X-ray Bulge'', seen most clearly in the southern Galactic
  Hemisphere, and stretching over the Galactic longitude range, $l$,
  from $|l| \la 30^\circ$ and down to approximately $-15^\circ$ in
  Galactic latitude.  This paper asks whether the X-ray bulge in the
  southern Galactic Hemisphere can be explained with a combined
  thermal and cosmic-ray driven wind.
  \label{r45Intensity}}
\end{center}
\end{figure*}

The Milky Way does exhibit clues that it might drive a large-scale
wind.  The first of these is an enhancement in the diffuse soft X-ray
emission, stretching over the longitude range $-20^\circ \la l \la
35^\circ$ with an emission scale height in the southern Galactic
hemisphere of $b \sim -17^\circ$ (see Fig.~\ref{r45Intensity}).  This
emission was first noted by \citet{SnowdenEtAl95}, who modeled it with
an isothermal plasma with a temperature $T = 4 \times 10^6$~K, a
midplane electron density of $n_{\rm e, midplane} \sim 3.5 \times
10^{-3}$~cm$^{-3}$, and a midplane thermal pressure of $P_{\rm g,
midplane}/k \sim 2.8 \times 10^4$~cm$^{-3}$~K.  At approximately the
same time, \citet{BreitschwerdtSchmutzler94} suggested that the
average \textit{all-sky} X-ray emission (not only that emission in the
region defined above) in all \textit{ROSAT} bands might be explained
by delayed-recombination in a large-scale cosmic-ray and thermally
driven wind \citep[see also][]{BreitschwerdtSchmutzler99}.

Later, \citet{AlmyEtAl00} used intervening absorption to show that at
least half of the central, enhanced X-ray emission lies more than
2~kpc from the sun \citep[see also][]{ParkEtAl97, ParkEtAl98}.  Since
that measurement was made in the Galactic plane, where the absorption
is strongest, it was inferred that most of the emission observed at
higher latitudes lies beyond that 2~kpc distance.  \citet{AlmyEtAl00}
also improved on previous modeling efforts: that work presents a model
of the emission due to a static polytropic gas (with $\gamma = 5/3$),
and very importantly, includes the effects of known background
components, such as the stellar background, extragalactic background,
and an additional isotropic background (to fit high-latitude
emission).  For comparison, their model had a central temperature of
$T_0 = 8.2 \times 10^{6}$~K, a central electron density of $n_{\rm e}
= 1.1 \times 10^{-2}$~cm$^{-3}$, and a central pressure of $P_{\rm
g,0} = 1.8 \times 10^5$~cm$^{-3}$~K.

We will compare our results with this static polytrope model to
investigate whether a wind model for this emission is feasible.

\subsection{Cosmic Ray Source Density}\label{CRdensity}

Another indicator of a Galactic wind comes from measurements of the
density of cosmic rays as a function of Galactocentric radius, $R$.
The source density of cosmic rays can be determined via $\gamma$-ray
emission: the production of $\gamma$-ray photons with energies
exceeding about 50~MeV is dominated by collisions of cosmic rays with
gas in the interstellar medium \citep{BlBlHe84}.  Since the galaxy is
largely transparent to such high-energy photons, the $\gamma$-ray
emissivity at those energies yields the cosmic ray source density.

If cosmic rays are produced in supernovae remnants, then since the
source density of supernovae remnants seems to increase with
decreasing $R$, the cosmic-ray source density should increase as well.
However, it has been known for some time \citep[e.g,][]{Bloemen89}
that the inferred cosmic-ray source density is relatively flat,
compared to the supernova density, as a function of $R$.

\begin{figure}[h]
\begin{center}
\includegraphics[width=6cm,angle=-90]{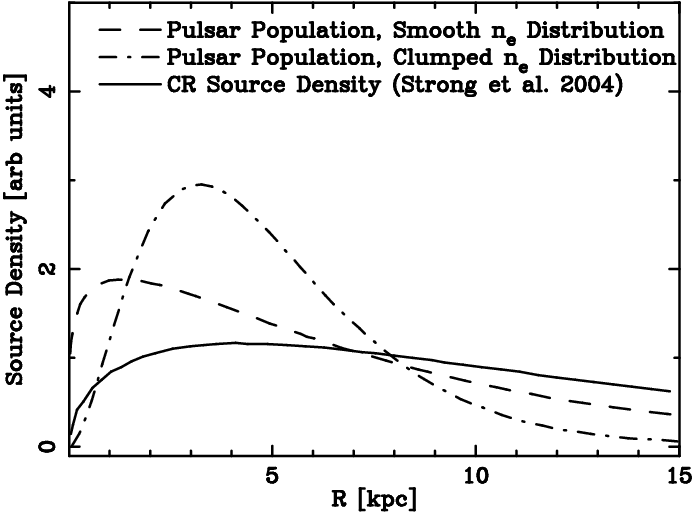}
\caption{Comparison of two different calculations of the pulsar
  population as a function of Galactocentric radius
  \citep{LorimerEtAl06} vs. the cosmic ray source density implied from
  the observed $\gamma$-ray emissivity \citep{StrongEtAl04}.  The two
  different curves for the pulsar distribution result from assuming a
  smooth distribution of $n_{\rm e}$ in the Galaxy \citep{LyMaTa85},
  or a clumped distribution, using \citet{CordesLazio02} and
  \citet{FaKa06}; for details, see \citet{LorimerEtAl06}.  The fact
  that the cosmic-ray distribution does not seem to follow the pulsar
  population has been known for some time \citep{Bloemen89}, but there
  is no consensus on the reason.  A cosmic-ray and thermal
  pressure-driven wind may help explain the cosmic-ray source
  population.
  \label{crSourceDensity}}
\end{center}
\end{figure}

There has, however, been some debate about whether supernovae remnants
are an accurate tracer \citep[since those surveys are subject to
various selection effects; see, e.g.,][and references
therein]{StrongEtAl04}.  Recent surveys of the pulsar population
\citep{LorimerEtAl06} also show that the pulsar source density
increases towards the center of the Galaxy, as shown in
Figure~\ref{crSourceDensity}.  This is true irrespective of the model
of how $n_{\rm e}$ varies in the disk, although the magnitude of the
pulsar population gradient with $R$ depends strongly on the $n_{\rm
e}$ model.  So, there remains a mismatch between the observed source
density of cosmic ray ``producers'' and the cosmic rays themselves.

It has already been pointed out that the observed slow rise in cosmic
rays may be due to a wind emerging from the disk, advecting cosmic
rays outwards \citep{BloemenEtAl93, BrDoVo02}.  In the case of
\citet{BloemenEtAl93}, a wind model was applied to the entire Galactic
disk; as a result, only a very slow wind was found to be compatible
with the inferred cosmic-ray source density.  In contrast,
\citet{BrDoVo02} applied their cosmic-ray and thermally driven wind
model, where the wind velocity varied as a function of radius and
height; they also took into account anisotropic diffusion.  With this
model, a small radial gradient in the cosmic ray source density could
be explained.

An alternate explanation for this slow change in the cosmic ray
population with $R$ was proposed by \citet{StrongEtAl04}, who found
that a radial variation in the $W_{\rm CO}$-to-$N(H_2)$ ratio by a
factor of 5 to 10 could explain the $\gamma$-ray observations.  In
this paper we primarily address the question of the origin of the
diffuse, soft X-ray background emission; we will, however, concentrate
on a large-scale wind model, keeping in mind its possible application
to the cosmic ray source density.

%Both the diffuse X-ray emission and the slow change in cosmic-ray
%population towards the Galactic center hint at the possibility of a
%large-scale wind driven by both thermal and cosmic-ray pressure.  We
%next outline a wind model including these components.

\section{A Cosmic Ray and Thermally Driven Wind Model}\label{modelDef}

%Many researchers have modeled thermal winds from the Galaxy, but it is
%important to point out that, as in the dynamics of the ISM
%\citep[e.g.,][]{Cox05}, cosmic rays may have a crucial role to play.
%This may be especially true for the two questions above, where we are
%interested in the thermal (soft X-ray) gas and the cosmic ray source
%density, both of which could be addressed by a two-component wind
%model.  So, we require a Galactic wind model that includes cosmic-ray
%pressure (exerted via \alfven waves) and thermal gas pressure.
%Therefore, we
%have built and will describe a Galactic wind model where the outflow
%is driven by a combination of cosmic-ray pressure (exerted via \alfven
%waves) and thermal gas pressure.

\defcitealias{BrMcVo1991}{BMV91}

To understand the observations outlined above, we must address both
the thermal (soft X-ray) gas and the cosmic ray source density.
Therefore, in investigating the possibility of an outflow, we require
a Galactic wind model that includes cosmic-ray pressure (exerted via
\alfven waves) and thermal gas pressure.  We outline such a model in
this section.

Our chief motivation here is to fit the wind model to the observed
large-scale soft X-ray emission, and as such, we require a relatively
simple model that can be computed quickly to compare with the
observations, but one that takes into account the physics of
cosmic-ray interactions with the thermal gas.  In addition, we are
interested in building intuition into the differences between pure
thermally-driven winds and winds with a significant cosmic-ray
component (see \S~\ref{ParameterSurvey}).  We are not the first to
address this; a very suitable model has already been developed by
\citet[][hereafter BMV91]{BrMcVo1991} and further advanced in later
papers \citep{BrMcVo1993,Zirakashvili1996,PtuskinEtAl1997,
BreitschwerdtSchmutzler99,BrDoVo02}.  This work had built on previous
analyses of the possibility of cosmic-ray driven winds
\citep{Ipavich1975, BVM87}.  In what follows, we explain this 1D,
semi-analytic wind model in detail, with particular attention to the
differences between our model and that of \citetalias{BrMcVo1991}.

\begin{figure}[h]
\begin{center}
\includegraphics[width=8cm, viewport=70 0 700 200, clip]{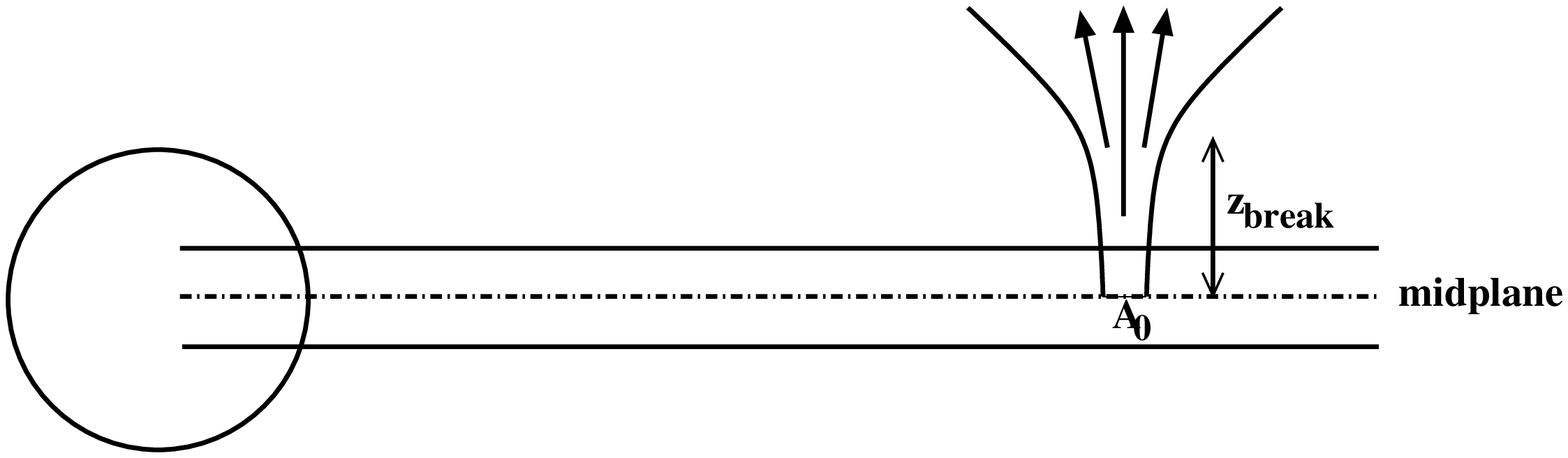} 
\caption{The geometry of a flow tube for the 1D wind model.  The wind
  starts at the Galactic midplane with cross-sectional area $A_0$,
  which is roughly constant up to $z \sim z_{\rm break}$, after which
  $A \propto z^{\alpha}$. \label{crWindTubeDiagram}}
\end{center}
\end{figure}

This 1D model must assume a particular geometric cross-section for the
wind; we adopt a flared-cylinder geometry shown in
Figure~\ref{crWindTubeDiagram}, as was adopted in \citetalias{BrMcVo1991} and used in
studies of coronal holes on the Sun \citep[e.g.,][]{KoppHolzer76}.  In
this geometry, the wind flows along a tube of approximately constant
cylindrical cross section up to a height $z \sim z_{\rm break}$, after
which the area increases as $z^{\alpha}$:

\begin{equation}
A(z) = A_0 \left[ 1 + \left( \frac{z}{z_{\rm break}} \right)^\alpha
  \right]. \label{areaEquation}
\end{equation}

We envision that this geometry is set by approximate pressure
confinement in the plane of the galaxy up to $z_{\rm break}$; above
that height, the area diverges spherically (thus $\alpha = 2$).
Unlike \citetalias{BrMcVo1991}, we allow $z_{\rm break}$ to change to
find the best-fit wind model.  We do not `freeze' this parameter, as
we have no \textit{a priori} value for $z_{\rm break}$ for this
particular wind geometry (which, in contrast to
\citetalias{BrMcVo1991}, covers only a part of the Galactic disk, and
so does not require a scale-height comparable to the scale-length of
the star-formation disk for an approximately spherical geometry, for
instance).  Thus, $z_{\rm break}$ is a fit parameter for our model, in
the context of the simplified geometry that we are assuming.
%Unlike \citetalias{BrMcVo1991}, we allow both $z_{\rm break}$ and $\alpha$ to change; we
%investigate their impact on the wind solutions in \S\S\ref{compareToObs}
%and \ref{ParameterSurvey}.  
Also unlike \citetalias{BrMcVo1991}, we assume that the wind is launched from the
Galactic midplane; this will be addressed in detail when the
interaction of cosmic rays, magnetic fields, and thermal gas is
discussed, below.

With this geometry, and assuming no additional mass loading of the
outflow, the wind has a simple equation of mass conservation:
\begin{equation}
\frac{d}{dz} (\rho v A) = 0,\label{conservationEq}
\end{equation}
where $z$ is the height above the Galactic midplane, $\rho$ is the
density in the wind and $v$ is the wind velocity.

Next, how is the wind driven along this flow tube?  We wish to
consider the possibility that cosmic-ray pressure, \alfven wave
pressure, and thermal pressure are all important components in driving
a Galactic wind.  At first glance, this may not seem productive, as
cosmic rays seldom interact directly with any particle in the galaxy:
the probability of any cosmic ray particle colliding with matter in
the Galaxy in their lifetime is of order $10^{-4}$ \citep[see,
e.g.,][]{Kulsrud05}.  However, cosmic rays are observed to have a very
small anisotropy (about 1 part in $10^4$), which seems at odds with
this low collision rate.  This small anisotropy is explained by
pitch-angle scattering of cosmic rays by \alfven waves in the ISM. It
has been shown that the cosmic rays which supply most of the pressure
($E\la 100$~GeV) can generate these waves via the ``streaming
instability'' \citep[e.g.,][]{Wentzel68, KulsrudPearce69,
KulsrudCesarsky71}.  This instability amplifies waves with wavelength
of order the cosmic ray gyroradius when the bulk velocity of cosmic
rays along the fieldlines exceeds the local \alfven speed. If the
cosmic ray mean free path is much shorter than global lengthscales in
the problem, the cosmic rays can be described as a fluid which moves
down its pressure gradient at velocity $v_A$ relative to the thermal
gas, while transferring momentum and energy to the waves, which in
turn transmits them to the thermal gas \citep[e.g.,][]{Skilling75,
DruryVoelk81, McKenzieWebb84}. We adopt that picture here.

In computing the interaction of cosmic rays, \alfven waves, and
thermal gas, we treat the cosmic rays as an ultra-relativistic
polytropic gas with $\gamma_c = 4/3$.  (This is an approximation, as
$\gamma_c$ is not exactly $4/3$; see \citeauthor{EnsslinEtAl06},
\citeyear{EnsslinEtAl06}.)  Meanwhile, we treat the thermal gas as
having a polytropic index, $\gamma_g$, of 5/3.  Much as in
\citetalias{BrMcVo1991}, we derive equations for the change in gas and
cosmic-ray pressure with height in the wind:
\begin{eqnarray}
\frac{dP_g}{dz}& = &\left( c_g^2 - 
               \gamma_c (\gamma_g-1) \frac{P_c}{\rho} \frac{1}{M_A}
               \frac{M_A + \frac{1}{2}}{M_A + 1} \right) \frac{d\rho}{dz},
               \label{gasPressureEq} \\
{\rm and}~~\frac{dP_c}{dz} & = & \frac{\gamma_c P_c}{\rho} \frac{M_A +
               \frac{1}{2}}{M_A + 1} \frac{d\rho}{dz},
               \label{crPressureEq}
%\frac{dP_w}{dz} & = & \frac{3 M_A + 1}{2 (M_A + 1)} \frac{P_w}{\rho}
%               \frac{d\rho}{dz} \label{wavePressureEq}
\end{eqnarray}
where $P_g$ and $P_c$ represents the gas and cosmic-ray pressure,
$c_g$ gives the sound speed in the gas, and $M_A = v/v_A$ is the
\alfven Mach number for the wind.  In the gas pressure equation,
Equation~\ref{gasPressureEq}, the first term in parentheses simply
relates the change in gas pressure to the change in density as the gas
accelerates and expands in the flow tube.

The second term in the parenthesis of Equation~\ref{gasPressureEq}
represents the coupling of cosmic-ray generated \alfven waves to the
gas; that term gives the heat input to the gas from the damping of
those cosmic-ray generated waves.  These waves represent the dominant
coupling between the cosmic rays and the thermal gas; this process
heats the gas despite the drop in density with height (hence the
negative sign for this term).  As the cosmic-ray generated \alfven
waves are immediately damped, they do not add to the wave pressure,
and hence we do not follow their evolution. In our models, the wave
pressure at the base of the wind is set to zero, reflecting the small
wave energy density in the Galactic plane ($\delta B/B\sim 10^{-3}$
from \citeauthor{KulsrudPearce69}, \citeyear{KulsrudPearce69}, where
$\delta B$ represents the \alfven wave perturbation to the large-scale
magnetic field, $B$).  Any energy transfered from the cosmic rays to
the waves is immediately input to the gas, as in
Equation~\ref{gasPressureEq} above.  The inclusion of this immediate
wave damping in all of the models presented here is an important
difference between this work and most of the models in
\citetalias{BrMcVo1991}; we also note that wave damping was considered
in much more detail in the later papers of \citet{Zirakashvili1996}
and \citet{PtuskinEtAl1997}.

This immediate damping of the cosmic-ray generated \alfven waves is
important for two reasons.  First, left unchecked the wave pressure
can easily grow to such magnitudes that the ratio of the perturbed
magnetic field to the large-scale magnetic field, $\delta B/B$,
exceeds unity.  In this limit, the derivation of the above equations
becomes suspect, as the system becomes nonlinear.

The second objection to unlimited \alfven wave growth is the presence
of rapid damping mechanisms in the ISM.  In particular, non-linear
Landau damping \citep{Kulsrud05} will quickly remove energy from the
waves.  The rate of non-linear Landau damping is of order
\begin{equation}
\gamma_{\rm NL} = \frac{\sqrt{\pi}}{4} \sqrt{\beta} \frac{ \delta B
  v_A \Omega}{B v_i} \sim \frac{1}{4} \sqrt{\frac{\pi}{2}}
  \left(\frac{\delta B}{B}\right)^2 \Omega_i
\end{equation}
where $\beta$ is the ratio of the gas pressure to magnetic pressure,
$(8 \pi P_g)/B^2$, $\Omega_i$ is the ion cyclotron frequency,
and $v_i$ is the ion thermal velocity.  For typical values for our
wind models at $z \sim 2$~kpc, ($B = 7.2 \mu$G, $\delta B/B \sim
10^{-3}$), we find $\gamma_{\rm NL} \sim 2 \times
10^{-8}$~s$^{-1}$, or $\tau_{\rm NL} \sim 5 \times 10^7$~sec $=
1.6$~yr.  We can compare this to the advection timescale, $1/\left((v
+ v_A) \cdot \nabla P_c/P_c \right)$, which for a typical wind model
is of order $10^{14}$~s, or $4.3 \times 10^6$~years.  From this
comparison, we can see that the damping is local, since the timescale
for non-linear Landau damping is much smaller than the advection
timescale.  Therefore, the local damping not only allows the
quasilinear equations to be applicable throughout the wind, but is
physically quite plausible, given the above small damping timescale.

(We note that these arguments were well-known to
\citetalias{BrMcVo1991}, but that the wind models calculated there did
not develop large $\delta B/B$ inside the critical point, so such
immediate damping was not important in those models [Breitschwerdt
2007, personal communication].)

We briefly consider the effect of \alfven waves generated by other
sources. Cosmic ray streaming is only one source of MHD turbulence in
the ISM, and, at large scales, by no means the dominant one. If the
turbulence which couples the cosmic rays to the gas were not primarily
due to the streaming instability, the model would be substantially
modified. The \alfven speed which appears in the model should be
thought of as the mean velocity of the waves in the rest frame of the
thermal gas; if the waves were isotropic, this speed would be
zero. The cosmic rays would simply behave as a polytropic fluid with
adiabatic index $\gamma_c$ and there would be no frictional heating of
the thermal gas.  This is not a significant issue, however, because of
the anisotropy in the turbulent cascade seen in both experiments and
theoretical studies of MHD turbulence \citep[see, e.g.][]{Shebalin83,
GS95, ChoVishniac00, Milano01}, which has also been invoked in studies
of turbulence in supernovae as well \citep{PtuskinZirakashvili03}.
The turbulence is initially presumably excited at scales of several
parsecs or more, far above the cosmic ray gyroradius scale of
10$^{12}$-10$^{13}$~cm, and cascades down to the gyroradius scale
through nonlinear wave-wave interactions.  In order to interact, these
waves must be oppositely directed along the magnetic field.  By
momentum and energy conservation, interactions between such
oppositely-moving waves yields resultant waves in which the component
of momentum along the magnetic field line has not increased.  However,
the perpendicular component can increase as a result of the
interaction, yielding an anisotropy in $\textbf{k}$-space
\citep{Shebalin83}.  Thus, at wavenumbers much greater than the
driving scale, the perpendicular wavenumber $k_{\perp}$ much exceeds
$k_{\parallel}$ \citep{GS95,ChoVishniac00}. More intuitively, for
motions on smaller scales and commensurately smaller energies, the
turbulent motions cannot bend magnetic field lines, and the energy is
tranferred to motions parallel to the magnetic field, resulting in
elongated eddies \citep{Shebalin83, Lazarian06}.

So how does such an anistropic eddy affect cosmic ray scattering?
When $k_{\perp}/k_{\parallel}\gg 1$, cosmic rays scatter much less
efficiently; this is because, when $k_{\perp} >> k_{\parallel}$,
individual cosmic rays sample many uncorrelated perturbations in
$k_{\perp}$ in each gyro-period that effectively cancel out
\citep{Chandran2000,Lazarian06}. Thus, the background Galactic
turbulent cascade has little effect on the cosmic rays compared to the
waves they generate themselves.  We note, however, that fast-mode
waves in the ISM may perhaps be important in scattering cosmic rays
\citep{YanLazarian03, YanLazarian05}.  For this work, however, we
retain the model of quickly damped, cosmic-ray generated \alfven waves
in Equation~\ref{gasPressureEq}.

Given the importance of gas heating, one may wonder if perhaps a loss
term, such as radiative cooling, is also important.  We have
determined that the radiative cooling is $\sim 1\%$ of the total power
in the wind, and so is unimportant for the wind models presented here.
However, we note that for very low velocity winds, the dynamical
timescale may exceed the radiative cooling time, resulting in
significant cooling.

We also ignore cosmic ray diffusion and thermal conductivity.
Diffusion is important towards the base of the wind
\citep{BrMcVo1993}; thermal conductivity may be important there as
well: our calculations show conductivity to be important for $z \la
350$~pc in the wind (below that height, the energy input from
conduction dominates adiabatic cooling and heating via wave-damping).
This is certainly significant, and both effects need to be considered
in a more detailed wind model (see \S\ref{Conclusions}).

Given Equations \ref{conservationEq} to \ref{crPressureEq}, and our
assumptions about the coupling of cosmic rays and thermal gas, the
above pressure relations are then coupled together in the wind
equation, which in its simplest form is
\begin{equation}
\rho v \frac{dv}{dz} + c_*^2 \frac{d\rho}{dz} = -\rho g, \label{windEquation}
\end{equation}
where $c_*$, the ``composite sound speed'' \citepalias[see][]{BrMcVo1991}, is given by
\begin{equation}
c_*^2 = \frac{d(P_g + P_c)}{d\rho}
\end{equation}
and where the gravitational acceleration, $g$, is defined by a
three-component (bulge, disk, and halo) model given in \citetalias{BrMcVo1991}.  We have
compared this gravitational potential model to the more recent work of
\citet{DehnenBinney98}, and found that, for $z > 200 $~pc, this newer
model gives a lower gravitational acceleration (by $\sim 10\%$ to at
most $\sim 60\%$) than the simple model of \citetalias{BrMcVo1991}; closer to the disk,
the model of \citet{DehnenBinney98} yields a higher gravitational
acceleration by $\sim 40\%$.  With the reduced calculational
requirements of the simpler potential (especially important near the
critical point of the wind), and the knowledge that the simpler model
largely over-estimates the gravitation potential where the wind is
accelerating, we choose the simpler model as a conservative
approximation; using the more realistic potential, launching a wind
from the Milky Way should be somewhat easier than shown here.

We solve the wind equation (Eq.~\ref{windEquation}) in the 1D flow
tube defined by the area law (Eq.~\ref{areaEquation}).  While
integrating, we use the simple magnetic flux conservation law,
$d(BA)/dz = 0$.  The integration is carried out using the
\texttt{dlsode} routine in ODEPACK \citep{Hindmarsh83}.

\subsection{Initial Conditions}\label{ICs}

The integrations require that $P_g$, $P_c$, $\rho$, $R$, and $B$ all
be specified at the base of the wind.  The value of $v$ is not given
at the base of the wind, as its value is set by requiring that the
integration pass through the critical point (see~\S~\ref{cpHandling}).
Fiducial values for these parameters are given in
Table~\ref{initParams}, and discussed in more detail below.

%% change!  initial conditions from Bob's comment about gas pressure
\begin{deluxetable}{lrr}
%\tabletypesize{}
%\tablenum{1}
%\tablecolumns{3}
\tablecaption{Initial (Midplane) Wind Parameters at $R = 3.5~kpc$\label{initParams}}
\tablehead{\colhead{Parameter} & \colhead{Value} & \colhead{Fixed?}}
\startdata
$R_0$ Range [Galactocentric] &  $1.5$ to $4.5$~kpc & Fixed  \\
$z_0$ & 1~pc & Fixed \\
$n_0$ & $1.8 \times 10^{-2}$~cm$^{-3}$ & Varied  \\
$P_{\rm g,0}/k_{\rm B}$ & $2.6 \times 10^4$~K~cm$^{-3}$ & Varied\\
$P_{\rm c,0}/k_{\rm B}$ & $2.2 \times 10^4$~K~cm$^{-3}$ & Fixed  \\
%$P_{\rm w,0}/k_{\rm B}$ & $1.7 \times 10^2$~K~cm$^{-3}$ & Fixed  \\
$B_0$ & $7.8~\mu$G & Fixed  \\
$z_{\rm break}$ & $4.5$~kpc & Varied   \\ 
$\alpha$ & $2.0$ & Fixed 
\enddata
\end{deluxetable}

\begin{figure}[h]
\begin{center}
\includegraphics[width=8cm, viewport=70 0 700 200, clip]{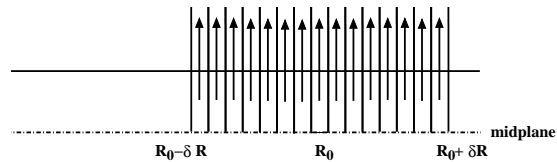} 
\caption{Near the Galactic midplane, the flowtubes are vertical,
  allowing the tubes to be assembled in a simple, large-scale wind
  model.  In this way, our 1D wind models are assembled into a
  large-scale wind to compare to the observed diffuse soft X-ray
  emission.\label{crWindTubeConcat}}
\end{center}
\end{figure}

The Galactocentric launching radius is set by the observed geometry of
the diffuse X-ray emission.  If the X-ray emission is roughly centered
on Sgr~A$^*$ \citep{AlmyEtAl00}, the extent of the emission out to
longitudes of $\sim 30^\circ$ implies a scale for the outer edge of
the wind at $R \sim 4.5$~kpc (assuming that the Sun is located at
$\sim8$~kpc).  The inner edge of the wind is set by the difficulty in
launching from the central Galactic potential; for ISM-like launching
conditions, we have found that models with $R \la 1.5~$kpc cannot
escape the Galaxy.  Thus, we set the geometry of our wind to be a
``thick curtain'', launched over Galactocentric radii of 1.5 to
4.5~kpc.  To run a fiducial model, we set the wind flowtube at
3.5~kpc.  More complicated simulations (where separate winds are
launched over the complete range or radii) have been run, and show
differences in X-ray emission of only $\sim 10\%$.  Therefore, the
wind models here are simply run at $R \sim 3.5$~kpc, and that wind is
then applied (or ``replicated'') to the range of wind radii (1.5 to
4.5~kpc) and over $2\pi$ in azimuth, to fit the observed diffuse X-ray
emission.  This geometry is shown in Figure~\ref{crWindTubeConcat},
below.

This is, of course, a simplification, but one that allows quick
calculations of the wind's X-ray emission and comparison to
observations (\S~\ref{compareToObs}), and allows surveys of large
areas of parameter space for building physical intuition
(\S~\ref{ParameterSurvey}) about galactic winds.  It is important to
note that, of course, at high latitudes (where $z \ga z_{\rm break}$),
each tube flares outward; with such tubes placed next to each other,
they will strongly overlap for $z \ga z_{\rm break}$.  This could be a
potential problem, but as we will see later (in \S\ref{fitObs}), the
X-ray emission is explained with $z_{\rm break}$ being more than a
factor of two larger than the emission scale-height, so this will not
impact our predictions of the X-ray emission.  It would, however, be
important for observational tests at higher latitudes.

We must also choose the initial height from which to launch the wind.
Again, we differ from \citetalias{BrMcVo1991}, and choose to launch
the wind from the midplane of the Galaxy, with $z_0 = 1$~pc.
\citetalias{BrMcVo1991} choose to launch from $z_0 = 1$~kpc due to
concerns about (1) ion-neutral friction due to partially neutral gas
for $z < 1$~kpc, and (2) an isotropic \alfven wave field (not
generated by the streaming instability of the cosmic rays).  While
those are all important considerations for the generic ISM, we
hypothesize that the hot, ionized medium from which these winds are
launched is largely free of neutrals, is dominated by heating and
cosmic rays due to nearby supernovae, and that cosmic-ray generated
waves dominate the scattering process (see \S\ref{modelDef}).  Also,
for meaningful comparison with observations, we must model the wind
below $z \sim 1$~kpc, as the scale height of the diffuse soft X-ray
emission is $z \sim 2$~kpc \citep{SnowdenEtAl97}.

In order to fit the wind models to the observed diffuse Galactic X-ray
emission, the variables $P_g$, $\rho$, and $z_{\rm break}$ will be
left to ``float'' (hence allowing the temperature and emissivity of
the gas the possibility of matching the observations).  But for
comparison, it is helpful to consider what prior observational
constraints we can place on these values.  Unfortunately, it is
difficult to constrain the thermal pressure of the hot ionized
component of the ISM at $R = 3.5$~kpc, or indeed, anywhere in the
Galactic disk \citep[see][for a summary]{Cox05}.  We estimate that at
the Sun's position in the Galaxy, $P_{g,\odot}/k_{\rm B} \sim
7000$~cm$^{-3}$~K \citep[][in his \S4.2]{Cox05}.  This wind is
launched much closer to the Galactic center, so that value must be
extrapolated to $R = 3.5$~kpc.  \citet{WolfireEtAl03} show that the
radial pressure scale-length is of order 5.5~kpc (notably, for the
neutral gas component), so to estimate $P_{g,0}(R = 3.5~{\rm kpc})$,
we must multiply $P_{g,\odot}/k_{\rm B}$ by $e^1$.  Therefore,
$P_{g,0}/k_{\rm B} \sim 1.9 \times 10^4$~cm$^{-3}$~K, or $P_{g,0} \sim
2.6 \times 10^{-12}$~dyne~cm$^{-2}$.

Estimates for the gas density are also difficult for $R \sim 3.5$~kpc,
but for a simple estimate, if we take $T \sim 10^6$~K for the hot
ionized medium (that this wind would be launched from), and use the
above pressure estimate, we find $n_0 \sim 1.8 \times 10^{-2}$, or
$\rho \sim 1.9 \times 10^{-26}$~g~cm$^{-3}$.

The cosmic-ray density can be estimated from the synchrotron
emissivity; much of the work for this has already been collected by
\citet{Ferriere01}.  We used her Equations~10 and 11, first
duplicating her Figure~7 to check our own implementation; we then use
those verified equations to calculate both $P_c$ and $P_{\rm
magnetic}$ (and hence $B$).  We find $P_c(R = 3.5~{\rm kpc}, z = 0) =
3.1 \times 10^{-12}$~dyne~cm$^{-2}$ ($= 2.2 \times 10^4$~cm$^{-3}$~K).
Similarly, $P_{\rm magnetic} = 2.4 \times 10^{-12}$~dyne~cm$^{-2}$ ($=
1.7 \times 10^4$~cm$^{-3}$~K), which implies $B = 7.8 \mu$G.

%In estimating $P_w$ at the galactic midplane, we use (as \citetalias{BrMcVo1991}) the
%approximation that $\delta B/B \sim 0.1$, and since $P_w =
%(\frac{\delta B}{B})^2 P_{\rm magnetic}$, this yields $P_w = 2.4
%\times 10^{-14}$~dyne~cm$^{-2}$.

We start with $\alpha=2$ \citepalias[as in][]{BrMcVo1991}, but with $z_{\rm break} =
4.5$~kpc, as many of our early wind models preferred $z_{\rm break}$ of
that order, as opposed to $z_{\rm break} = 15$ in \citetalias{BrMcVo1991}.

\subsection{Integrating Through the Critical Point}\label{cpHandling}

Like many other similar wind models, this wind equation contains a
critical point (CP) where $v = c_*$.  Such critical-point equations
are usually solved by integrating from the singular point to the
boundaries (here at $z = 1$~pc and $z = 1$~Mpc).  The dependence of
the wind on cosmic-ray pressure and gas pressure mean that, in order
to follow this usual procedure, estimates of those pressures would
have to be known \textit{a~priori} at the critical point position.  We
choose \citepalias[as did][]{BrMcVo1991} to estimate the pressures at
the launching point of the wind, searching for a $v_0$ which defines
the wind that threads the critical point.  We checked our
implementation by recalculating models presented in Table~1 of
\citetalias{BrMcVo1991}.
%To complete the final wind integral, we integrate
%to near the critical point using double precision calculations, with
%relative errors limited to 10$^{-15}$ per integration step.  In the
%vicinity of the CP [$z \sim z_{\rm CP}(1 \pm 10^{-3})$], we shift to a
%lower precision integral (with relative precision of only 10$^{-10}$
%per integration step), utilize an approximation for $\left(
%\frac{du}{dz} \right)_{\rm CP}$ (as in \citetalias{BrMcVo1991}), and then after passing
%the critical point, shift back into full precision integration mode
%again.  This method was not strictly necessary for all runs (the
%integration code could, in many cases, integrate directly through the
%approximate critical point position), but implementing this mode made
%the code extremely robust over the more-than order-of-magnitude change
%in parameters that we will investigate.
%
%\subsection{Checking the Code}
%
%We checked our implementation by recalculating models presented in
%Table~1 of \citetalias{BrMcVo1991}.  Calculations with the lower-precision integrations
%near the critical point were checked vs. earlier full-precision
%calculations, and found to be identical within the numerical
%precision.

This concludes our definition of the cosmic-ray and thermally driven
wind model that we will compare to the soft X-ray data.  The next
section therefore returns to the X-ray data briefly mentioned in
\S\ref{Intro} and reports on our search for a wind model to fit those
observations.

\section{Comparing the Model to \textit{ROSAT} Observations}\label{compareToObs}

One of the main goals of this work is to investigate whether a wind
model can reproduce the diffuse soft X-ray background observations
most recently presented in \citet{SnowdenEtAl97}.  In this section, we
describe the algorithm we developed to calculate that emission.

\subsection{Calculating the Average Observed X-ray Emission}

We will compare the models against the longitude-averaged X-ray
emission as a function of latitude.  We retrieved the diffuse X-ray
emission maps from
{\small {\url{http://www.xray.mpe.mpg.de/rosat/survey/sxrb/12/fits.html}}}
at the ``X-Ray Astronomy'' page at the Max-Planck-Institut für
extraterrestrische Physik.  These maps are the full 12' resolution
maps; the emission maps are then averaged over the longitude range
from $-30^\circ < l < +30^\circ$, binned $1^{\circ}$ in latitude, and
plotted vs latitude for the southern Galactic hemisphere only.
%\footnote{The average emission in the plots shown in this paper
%is significantly ($\sim 20\%$) less than that presented in
%\citet{AlmyEtAl00}; we have determined that the averages in that paper
%only included data from $0^\circ < l < +30^\circ$.}  
The Northern Galactic hemisphere contains other features (such as the
North Polar Spur) that make model comparisons there much less clear.
Note again that there is a slight asymmetry in the observed X-ray
emission towards positive Galactic longitudes (towards the right on
Fig.~\ref{r45Intensity}); we are not modeling that asymmetry here.

The resultant longitude averaged, observed emission is shown by the
diamonds in Figure~\ref{compareWindToData-R4} (for the \textit{ROSAT}
R4 band, centered at approximately 0.65~keV) and
Figure~\ref{compareWindToData-R5} (for the R5 band, centered at
approximately 0.85~keV).  Error bars (from the original data files)
are given by the vertical lines within the diamond-indicated
data-points; those error bars assume only statistical errors in the
measured X-ray flux.  Systematic uncertainties in the foreground and
background components (particularly the stellar contribution) may be
larger.

We restrict ourselves to the R4 and R5 bands as the emission in those
bands comes primarily from oxygen emission lines; higher energy bands
(near 1.5~keV, for instance) may depend more strongly on metallicity,
as magnesium and silicon emission lines begin to dominate at those
energies.  In addition, at higher energies, the contribution from the
stellar background becomes much more prominent and that background is
not well understood (see \S\ref{backgrounds}).

\subsection{Calculating the Wind's X-ray Emission}

The wind model gives $n(z)$ and $T(z)$ along 1D streamlines.  As
mentioned in \S~\ref{ICs}, the observed large-scale emission is
simulated with this wind model by ``replicating'' the wind solution
both in radius (from 1.5 to 4.5~kpc) and in azimuth (see
Fig.~\ref{crWindTubeConcat}).

First, to calculate the emissivity per emission measure, the emission
codes ``ATOMDB'' and ``APEC'' \citep{SmithEtAl01} were used to
generate spectra for a range of temperatures\footnote{As noted on
their website (\url{http://cxc.harvard.edu/atomdb/}), APEC is not
complete below about 0.25~keV, but that will not greatly affect the
predictions here, since the relevant bands for this work are at higher
energies.} from $10^{5.6}$ to $10^{7.6}$~K in steps of $10^{0.1}$~K.
These codes use the metallicities of \citet{AG89}.  Each APEC-generated
spectrum was then folded through the \textit{ROSAT} response
matrix\footnote{We used a file \texttt{pspcc\_gain1\_256.rsp},
downloaded on Feb. 22, 2006 from the ``X-ray Background Tool'' on the
HEASARC website at
\url{http://heasarc.gsfc.nasa.gov/docs/tools.html}}, with the
responses of the individual channels summed into groups to represent
the \textit{ROSAT} passbands R4 and R5
\citep{SnowdenEtAl95}\footnote{NB: We found that the channel
boundaries reported in \citet{SnowdenEtAl95}, when used to sum up the
channel-by-channel responses to create the R4 and R5 band responses,
\textit{did not} recreate the response matrices plotted in
\citet{SnowdenEtAl97} and \citet{AlmyEtAl00}.  The only way to
reproduce the previous response matrices was to subtract 6 channels
from the channel boundaries in \citet{SnowdenEtAl97}.  In order to
compare the wind model to the data, it is essential that we use as
similar a response matrix as possible to that used in
\citet{SnowdenEtAl97}, so we apply this channel offset.}.
%The response matrix used was developed for modelling point sources; as
%such, its units are essentially cm$^2/(4 \pi$ st); to convert this to
%work with the diffuse X-ray maps, we convert steradians to arcmin$^2$
%by dividing by $11.8 \times 10^6$~arcmin$^2$/st.  
Applying these general models to the wind, the resulting matrix of
emissivity per emission measure vs temperature for each band is then
interpolated to calculate the emissivity at each point in the wind
model.  We find that the newer APEC-derived models yield a maximum of
$50\%$ more emission in the M-band (the combined \textit{ROSAT} R45
band) than the \citet{RaymondSmith77}-derived models of
\citet{AlmyEtAl00} near $T = 4 \times 10^6$~K, but the differences are
only of order $20\%$ near $T = 2 \times 10^6$~K.

The emission measure is calculated by simply summing $n_e^2\Delta l$
along lines-of-sight through the wind model.
%, but with the emission
%measure broken down as a function of temperature.  This is then
%multiplied, temperature bin by temperature bin, by the emissivity per
%unit emission measure.  The various temperature bins are then summed
%to give the total observed emission.  
This model emission is then corrected for absorption by applying (as a
foreground absorption screen) the $N_H$ data of
\citet{DickeyLockman90} for each line of sight.  The resultant wind
emission for both the R4 and R5 bands is shown as the dashed line in
Figures~\ref{compareWindToData-R4} and \ref{compareWindToData-R5},
respectively.

We note briefly that we have checked that the gas in these winds is in
equilibrium throughout the region where X-ray emission is important.
This has been verified with the non-equilibrium cooling code of
\citet{BBC01}. For instance, for the best-fit wind model presented in
\S\ref{fitObs}, we have found that non-equilibrium calculations yield
only $1\%$ differences in the population of fully-ionized oxygen
vs. equilibrium at $z = 2$~kpc, whereas for less ionized states of
oxygen, non-equilibrium calculations show $1\%$ deviations only beyond
$z\sim4.5$~kpc.  Since we are interested in the X-ray emission at $z
\la 2$~kpc, we retain simple collisional-equilibrium models.  The
previous Galactic outflow model of \citet{BreitschwerdtSchmutzler99},
also of a hybrid cosmic-ray and thermal-gas pressure driven wind,
relied on non-equilibrium effects \citep{BreitschwerdtSchmutzler94} to
model the full-sky \textit{ROSAT} emission in the all of the observed
bands; the present model concentrates solely on the excess towards the
Galactic Center, and for reproducing these observations and for this
wind model, we find that collisional equilibrium dominates.
Non-equilibrium effects may certainly be important at very large
heights ($z \sim 20$~kpc) and correspondingly cool temperatures, but
we do not find those effects to be significant in reproducing the
\textit{ROSAT} R4 \& R5 emission.

\subsection{Calculating the Background Emission}\label{backgrounds}

As first considered by \citet{AlmyEtAl00}, various background sources
also contribute emission to the soft, diffuse X-ray observations.
%We
%outline the calculation of each of those backgrounds in this section.
The stellar and extragalactic backgrounds listed below were used in
\citet{AlmyEtAl00}, and were both folded through both the Galactic
$N_H$ map and through the \textit{ROSAT} response matrix.

%\subsubsection{Extragalactic Background Emission}

The extragalactic emission is calculated using the power-law given in
\citet{HasingerEtAl93}: $7.8~(E/1 {\rm
keV})^{0.96}$~keV~cm$^{-2}$~s$^{-1}$~sr$^{-1}$~keV$^{-1}$ for $E <
1$~keV.  
%This is then absorbed by Galactic $N_H$ \citep[again,
%using][]{DickeyLockman90}, and then folded through the \textit{ROSAT}
%response matrix and averaged over longitude.  
This averaged extragalactic background is shown as the
dot-dot-dot-dashed line in Figures~\ref{compareWindToData-R4} and
\ref{compareWindToData-R5}.

%\subsubsection{Galactic Unresolved Stellar Emission}\label{stellarBkgd}

The stellar background used is the model of \citet{SchmittSnowden90}.
%,
%again folded through the Galactic absorption model and through the
%\textit{ROSAT} response matrix.  
This background model is shown as the long-dashed line in
Figures~\ref{compareWindToData-R4} and \ref{compareWindToData-R5}.
There are significant uncertainties with this stellar background
model; \citet{KashyapEtAl92} claim that there are errors of order a
factor of three in \citet{SchmittSnowden90} due to the assumed
luminosity function; unfortunately, however, \citet{KashyapEtAl92} do
not model the stellar emission towards the Galactic Center.  We retain
the stellar model of \citet{SchmittSnowden90} because it addresses the
stellar background in that area.  The uncertainty in this background
represents another reason for the present work to not address the
1.5~keV emission, which would be even more strongly contaminated by
coronal emission; emission in the R4 and R5 bands is not as strongly
affected by the stellar background.

%\subsubsection{Isotropic Emission}

All of the above extragalactic and background emission, added
together, do not fully account for emission at high latitudes ($b <
-60^\circ$).  The wind model emission is also insignificant at this
height.  To account for this excess emission, we add in an isotropic
component (the dotted line in Figs.~\ref{compareWindToData-R4} and
\ref{compareWindToData-R5}).  This emission may be due to
poorly-understood disk and halo emission that provides a rather
uniform and apparently thermal-emission background at high latitudes
and towards the Galactic anticenter.  This component is simply fitted
(for each model fit to the data) to account for the emission in the
latitude range $-90^\circ < b < -60^\circ$.

\subsubsection{Refitting the Static Polytrope Model}

The best-fit wind model will be found by comparison with the
observations, below (\S\ref{fitObs}), but it is also instructive to
compare the wind model to the static polytrope model of
\citet{AlmyEtAl00}.  Having updated several steps in the data and
model analysis procedures, we re-fit the polytrope model to the data
to ensure that both models are given fair consideration.  We therefore
adapted the wind-model routines and parameter-search codes to produce
new polytrope models and find the best polytrope fit, again using
$\gamma=5/3$ as in \citet{AlmyEtAl00}.  We produced a grid of 101
$\times$ 101 polytrope models, stepping logarithmically between
$P_0/k_{\rm B}$ from $1.8 \times 10^4$ to $1.8 \times
10^6$~cm$^{-3}$~K and through $k$ values of $1.45 \times 10^{31}$ to
$1.45 \times 10^{33}$~cm$^4$~g$^{-2/3}$~s$^{-2}$.  This parameter
survey confirmed that the \citet{AlmyEtAl00} values for those
polytrope constants remains the best-fit: we find $P_0/k_{\rm B} = 1.8
\times 10^5$~cm$^{-3}$~K and $k = 1.45 \times
10^{32}$~cm$^4$~g$^{-2/3}$~s$^{-2}$.  This model, added to the
background sources already considered, yields the polytrope models
shown by the dot-dashed lines in Figures~\ref{compareWindToData-R4} and
\ref{compareWindToData-R5}.

\subsection{Fitting the Wind to the Observed Emission}\label{fitObs}

The longitude-averaged wind model's X-ray emission is added to the
various background components, and the sum is given as the solid black
line in Figures~\ref{compareWindToData-R4} and
\ref{compareWindToData-R5}.  For each attempted model fit (see below),
$\chi^2$ is calculated by considering the full range of latitudes
plotted, weighted by the errors shown in the data points.  The reduced
$\chi^2$ values are not close to unity, but we retain $\chi^2$ as a
relative figure-of-merit for comparing a wide range of models.

To find the best-fit model, an initial parameter survey was carried
out over two orders of magnitude in $\rho_0$ and $P_{g,0}$ (centered
near the initial parameters given in Table~\ref{initParams}), and a
factor of 4 in $z_{\rm break}$; the parameter ranges of $\rho_0$ and
$P_{g,0}$ were gridded into 35 steps; while $z_{\rm break}$ was
gridded into 5 bins.  After this more general survey found the
approximate location of the best-fit model, higher resolution sampling
was carried out, leading to steps of $\sim 1.3\%$ in $\rho_0$ and
$P_{g,0}$, and $\sim 6\%$ steps in $z_{\rm break}$ near the $\chi^2$
minimum.  The best fit model parameters are given in
Table~\ref{bestFitParams} with X-ray emission shown in bands R4 and R5
in Figures~\ref{compareWindToData-R4} and \ref{compareWindToData-R5},
respectively.  The best-fit model's position in $\chi^2$ space is
fairly well constrained, as shown by the green ellipse in
Fig.~\ref{mDotPlot}.  Notably, at fixed $z_{\rm break}$, the fits for
the individual \textit{ROSAT} emission bands are \textit{very} close
together; the best-fit R5 model is identical to the joint best-fit R4
\& R5 model, and the best-fit R4 model differs by only $9\%$ in
$P_{g,0}$.

Interestingly, the $P_{g,0}$ value obtained is quite similar to the
ISM value that was initially estimated.  Granted, our initial, pre-fit
estimate of $P_{g,0}$ (in Table~\ref{initParams}) could have
uncertainties of at least a factor of two, but it is somewhat
satisfying that the wind's required pressure is relatively close to
the nominal ISM thermal pressure at the launch position.  Meanwhile,
the density at the base of the wind is $1/3$ of our estimate of the
ISM density.  

It is important to note also that $P_{c,0} \sim P_{g,0}$, and so in
this best-fit model, \textit{cosmic rays are an important component in
driving a wind from the Milky Way}.  In fact, in the best-fit wind
model, $P_{c,0}$ is very slightly greater than $P_{g,0}$ (but only by
$10\%$).  Of course, our value of $P_{c,0}$ is set from synchrotron
measurements \citep{Ferriere01}: the diffuse X-ray emission by itself
can place only fairly weak constraints on the cosmic-ray pressure, and
only then because the damping of cosmic-ray generated waves increases
the temperature of the gas slightly (in the best-fit model, $T_{\rm
max} \sim 1.15 T_0$), or leads to a wind that cannot escape the Milky
Way.  Judging from an increase of $\chi^2$ by $\sim 50\%$, the X-ray
data constrains $P_{c,0}$ to lie between $\sim10^4$~cm$^{-3}$~K and
$\sim6\times10^4$~cm$^{-3}$~K.  (For $n_0$ constant, the best fit
$P_{c,0}$ is $2.7~\times 10^4$~cm$^{-3}$~K, only $\sim~20\%$ away from
our assumed $P_{c,0}$.)

The parameter $z_{\rm break}$ is also important to fitting the
\textit{ROSAT} data.  Moving $z_{\rm break}$ directly impacts the
location of the critical point, and therefore the fall-off of X-ray
intensity with height.  Having no \textit{a priori} constraint for the
value for $z_{\rm break}$ for this geometry, we allowed the parameter
to float, finding the best fit value of $z_{\rm break} = 5.2$~kpc.
Again, judging from an increase in the minimum $\chi^2$ by $\sim
50\%$, $z_{\rm break}$ is approximately constrained to lie between
$4.1$~kpc and $6.5$~kpc.  To help guide intuition: as $z_{\rm break}$
decreases, the critical point position decreases, the mass outflow
rate increases, and the total energy required increases.

\begin{deluxetable}{lrl}
%\tabletypesize{}
%\tablenum{1}
%\tablecolumns{3}
\tablecaption{``Best-Fit'' Wind Parameters\label{bestFitParams}}
\tablehead{\colhead{Parameter} & \colhead{Value} & \colhead{Fixed?}}
\startdata
$P_{\rm g,0}/k_{\rm B}$ & $2.0 \times 10^4$~K~cm$^{-3}$ &
Varied\\ % 3.16e-12 dyne cm^-2
$n_0$ & $6.9 \times 10^{-3}$~cm$^{-3}$ & Varied  \\ % 5.35e-27 g/cm^3
$z_{\rm break}$ & $5.2$~kpc & Varied   \\ 
$R_0$ Range [Galactocentric] &  $1.5$ to $4.5$~kpc & Fixed  \\
$P_{\rm c,0}/k_{\rm B}$ & $2.2 \times 10^4$~K~cm$^{-3}$ & Fixed  \\
%$P_{\rm wave,0}/k_{\rm B}$ & $1.7 \times 10^2$~K~cm$^{-3}$ & Fixed  \\
$B_0$ & $7.8~\mu$G & Fixed  \\
$\alpha$ & $2.0$ & Fixed 
\enddata
\end{deluxetable}

\begin{figure}[h]
\begin{center}
\includegraphics[width=6.5cm,angle=90]{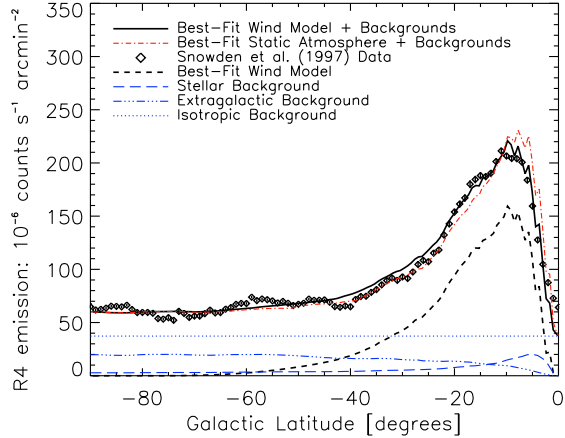} 
\caption{\textit{ROSAT} R4-band emission (centered on $\sim 0.65$~keV)
  from the best fit wind model (Table~\ref{bestFitParams}) compared to
  the longitude-averaged diffuse X-ray emission from
  \citet{SnowdenEtAl97}. The \textit{ROSAT} data points are plotted as
  diamonds, with vertical lines representing the error bars; the error
  bars are of very similar size to the plotting symbols.  In the R4
  band, the wind and static-polytrope models both fit the data
  reasonably well, although the $\chi^2$ for the wind is 2.1 times
  smaller than that for the static-polytrope. Still, systematic
  deviations dominate: $\chi_{\nu}^2 = 19.0$ for the best-fit wind
  model in the R4 band.\label{compareWindToData-R4}}
\end{center}
\end{figure}

\begin{figure}[h]
\begin{center}
\includegraphics[width=6.5cm,angle=90]{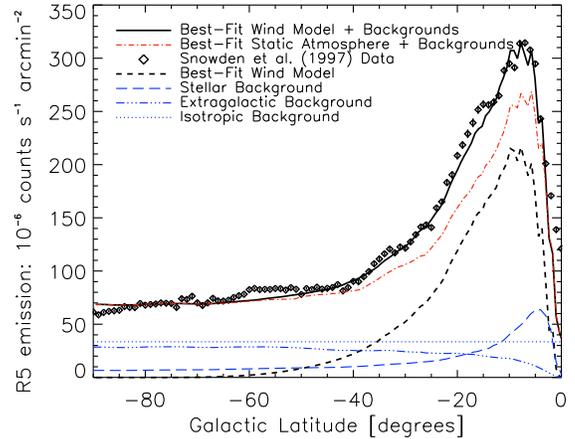} 
\caption{\textit{ROSAT} R5-band emission (centered on $\sim 0.85$~keV)
  from the best fit wind model (Table~\ref{bestFitParams}) compared to
  the longitude-averaged diffuse X-ray emission from
  \citet{SnowdenEtAl97}.  The \textit{ROSAT} data points are plotted
  as diamonds, with vertical lines representing the error bars; the
  error bars are of very similar size to the plotting symbols.  In the
  R5 band, the wind models fits much better than the static-polytrope
  model, with a difference of 2.3 in $\chi^2$.  As for band R4,
  $\chi_{\nu}^2$ is not near unity, with a value of 48.9; again,
  systematic deviations from the model dominate.
  \label{compareWindToData-R5}}
\end{center}
\end{figure}

The comparison between the wind model, static polytrope model, and
data shows that the wind is an improved fit to the data.  The $\chi^2$
values for bands R4 and R5 are smaller by a factor of 2.1 and 2.3 for
the wind models compared to the static polytrope model.  The $\chi^2$
values are themselves very high ($\sim 5.9 \times 10^3$ for the best
fit model), showing that there are still many deviations between the
data and wind model, but an examination of
Figures~\ref{compareWindToData-R4} and \ref{compareWindToData-R5} show
that this simple wind model is at least a reasonable fit to the data,
and therefore merits continued detailed consideration.

%\begin{figure*}
%\begin{center}
%\includegraphics[width=14cm,angle=90]{figures/windModelJointChiSqMap_color.ps} 
%\caption{$\log(\chi^2)$ space vs $n_0$-$P_{\rm g,0}$.  Note that the
%  best-fit model is well-constrained to a small area of parameter
%  space (both for the joint R4 and R5 bands, and for each band
%  separately).  Note that only successful wind models are plotted; the
%  shape of the boundary around the above contours will be described in
%  detail in \S\ref{ParameterSurvey}.
%  \label{chi2space}}
%\end{center}
%\end{figure*}

\subsection{The Best-Fit Wind Model}

\begin{figure}[h]
\begin{center}
\includegraphics[width=6.5cm,angle=-90]{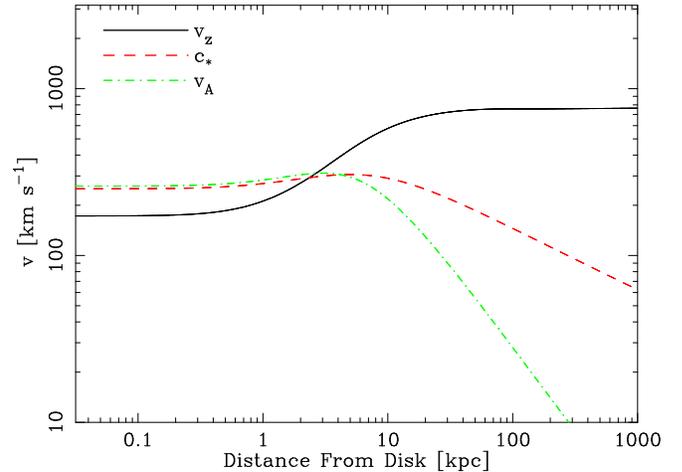} 
\caption{Velocity vs. height in a fiducial wind model.  The solid line
  represents the wind velocity, the dashed line represents $c_*$, the
  composite sound velocity, and the dot-dashed line shows the change
  in the \alfven velocity with height.  This velocity curve shows the
  rather standard increase in velocity of a pressure-driven wind,
  rising from the relatively low $v_0$, through the critical point at
  $v = c_*$, and accelerating on to $v \sim v_{\infty}$.
  \label{crWindVelPlot}}
\end{center}
\end{figure}

Before moving on to consider the range of models produced in our
parameter surveys, it will help to consider the best-fit model in some
detail.  In Figure~\ref{crWindVelPlot}, we show the trends in velocity
vs height for the best-fit wind model.  The solid line represents the
velocity curve for outflowing gas in the wind; like all thermal winds,
it starts at an initial velocity less than the sound speed,
accelerates through the critical point (where $v = c_*$) and
asymptotes at a velocity of order the escape velocity.  Interestingly,
in this model, the position of the critical point is $z_{\rm CP} =
$2.4~kpc, near the edge of the region of observed emission.  Also,
this wind requires an initial velocity of $v_0 = 173$~km~s$^{-1}$; in
comparison, at the base of the wind, $c_{s,0} = 198$~km~s$^{-1}$ (from
$T_0 = 2.9 \times 10^6$~K) while $c_{*,0} = 251$~km~s$^{-1}$.  The
terminal velocity of the wind is $v_{\infty} \sim 760$~km~s$^{-1}$.

%Also in this plot, the \alfven velocity and composite sound speed are
%shown for comparison.  Both drop with height due to the acceleration
%(and therefore drop in density) of the wind and the expansion of the
%flow-tube that the wind is embedded in.

\begin{figure}[h]
\begin{center}
\includegraphics[width=6.5cm,angle=-90]{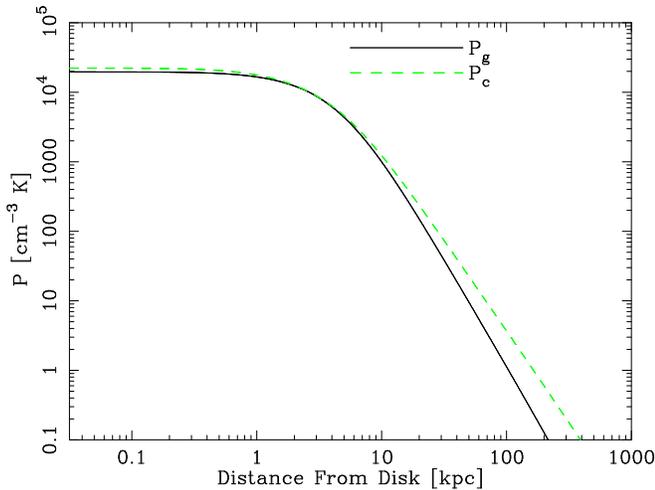} 
\caption{Pressure vs. height in a fiducial wind model.  The solid line
  shows the thermal pressure and the dashed line shows the cosmic-ray
  pressure.  Because $\gamma_c = 4/3$ and $\gamma_g = 5/3$ in our
  model, the cosmic-ray pressure drops off more slowly with height in
  the wind.  In addition, as cosmic-ray driving is somewhat less
  efficient than thermal driving (see \S\ref{cosmicRayDriving}),
  the thermal pressure is most significant at the base of the wind,
  whereas the cosmic-ray pressure is most significant at large (kpc)
  heights.  
  \label{crWindPressurePlot}} 
\end{center}
\end{figure}

Figure~\ref{crWindPressurePlot} shows how the gas and cosmic-ray
pressures compare, and how they each change with height in the wind.
The cosmic-ray pressure and gas pressure are nearly equal at the base
of the wind, but at large scales the cosmic-ray pressure drops off
less quickly than thermal pressure, as we have assumed that
$\gamma_{c} = 4/3$ and $\gamma_{g} = 5/3$ (for cosmic rays and gas,
respectively).  The large-scale importance of cosmic rays will be
investigated further, below.
%The wave pressure is negligible compared to the other
%pressure components.  This is in part due to the fact that all
%momentum transfered to \alfven waves from the cosmic rays is
%immediately damped to the gas; thus, the wave pressure is not allowed
%to build to levels comparable to the other pressures.

\begin{figure}[h]
\begin{center}
\includegraphics[width=6.5cm,angle=-90]{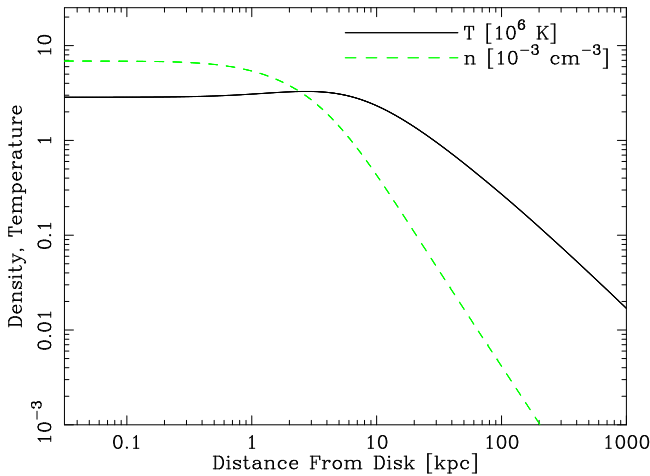} 
\caption{Normalized density and temperature vs. height in a fiducial
  wind model. The temperature is normalizes to $10^6$~K and the
  density is normalized to $10^{-3}$~cm$^{-3}$.  The temperature
  briefly increases to a peak at $z \sim 3$~kpc because of the damping
  of cosmic-ray derived \alfven waves.  The density, however, drops
  off quickly enough with height that the gas pressure remains
  monotonic (see Fig.~\ref{crWindPressurePlot}), as it must for the
  wind to be driven to large heights.
  \label{densityTemperaturePlot}}
\end{center}
\end{figure}

We present the changes in density and temperature with height in
Figure~\ref{densityTemperaturePlot}.  The temperature shows an
increase at around $z \sim 2-3$~kpc due to the damping of the
cosmic-ray generated waves.  The density simply falls off with height
as expected from mass conservation; the density drops off so quickly
that the increase in temperature due to wave damping does not yield an
increase in gas pressure at large height, although the gas pressure
does drop off more slowly with height than it would in the absence of
cosmic-ray wave damping.

The importance of this wind for the Galaxy is largely summed up in the
mass outflow rate: for this model, $\dot{M} =
2.1$~M$_{\odot}$~yr$^{-1}$.  This is a very large mass outflow rate,
but is in the same range as the inferred mass inflow rate to the
Galaxy of $\sim 1 - 10$~M$_{\odot}$~yr$^{-1}$ \citep{Bregman99}.  Even
with that infalling gas, this wind would have important implications
for metallicity gradients in the Milky Way.  However, we note that
while that mass outflow rate is well-constrained within the context of
this simple model, there are factors which we do not consider which
may significantly decrease the required mass outflow rate.  For
instance, clumping in the wind (see \S\ref{Conclusions}) could lower
the mass outflow rate required to supply the observed emission; the
``best-fit'' wind here (with constant density at the base) would
therefore represent a maximal mass-loss wind.

\subsection{Why Does the Wind Fit Better than a Static Atmosphere?}

Overall, one might expect the wind shown here to yield emission
similar to a static atmosphere, as the outflow is close to hydrostatic
equilibrium within the critical point.  This is true, but there is one
principal reason why the wind improves upon the previous static
atmosphere models.  The wind is defined to lie between 1.5 and
4.5~kpc: the outer radial bound does not affect the difference of fit
(since the longitude range of the survey data used here is limited
anyway), but the inner bound is important, and is physically
motivated: the Galactic potential there makes launching a wind
unlikely, except under extreme conditions.  (There could, of course,
still be X-ray emitting gas within the Galactic core representing a
fraction of the observed emission; the inclusion of such emission
would lower the required mass outflow rate in this outflow.)  In
contrast to the wind model, the static polytrope must, by definition,
fill the core of the Milky Way.  This leads to extremely high
temperatures (given the deep potential) and over-estimates of the
emission near the Galactic center in order to have approximately
correct emission temperatures elsewhere, although perhaps thermal
conduction could compensate for this \citep{AlmyEtAl00}.

Also, we note that the drop-off in temperature with distance in the
static polytrope model, and the necessity of the central very
high-temperature region, means that the temperature elsewhere in the
wind is, on average, somewhat smaller than required to explain the
emission.  This can be seen in Figure~\ref{compareWindToData-R5}.

Compared against the static-atmosphere model, then, the wind model is
the preferred fit for the diffuse soft X-ray data.  But how do such
wind models work as the fit parameters are varied?  These questions
are addressed in the following section.

\section{How Galactic Winds Change with Initial Parameters:
  Building Intuition}\label{ParameterSurvey}

The parameter surveys that yield a best-fit model are also quite
useful for building intuition and understanding of mixed cosmic-ray
and thermally-driven winds.  We present a few of the key results
below.  For each survey, we have varied the parameters around the
best-fit values for the Milky Way to understand how the outflow would
change character near the best-fit parameter values.

\subsection{Mass Outflow Rate}\label{mDotMap}

\begin{figure*}
\begin{center}
\includegraphics[width=10cm,angle=90]{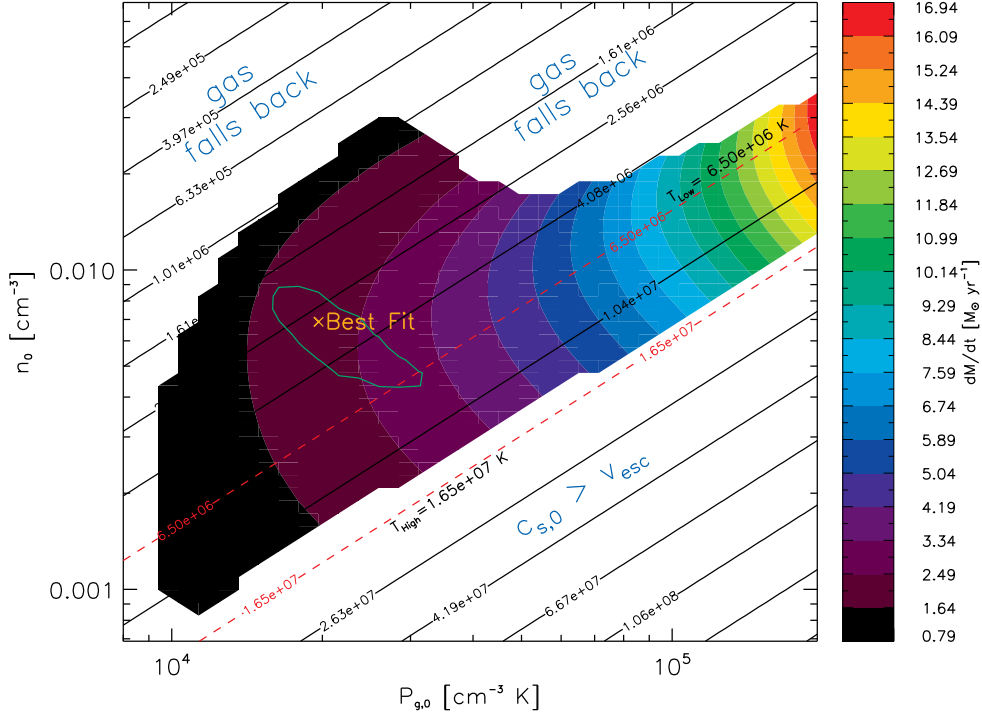} 
\caption{Color contours of mass outflow rate in units of
  $M_{\odot}$~yr$^{-1}$, with temperature (in K) represented by solid
  black contour lines, as a function of base density, $n_0$, and gas
  pressure, $P_{\rm g,0}$.  The shaded contour region shows those
  winds that pass through a critical point, as distinct from those
  regions where $T$ is too high ($c_{\rm s,0} > v_{\rm esc}$) and
  where $T$ is too low, and the gas falls back.  These regions of
  failed wind are approximately defined by either $T \ga T_{\rm high}$
  or $T \ga T_{\rm low}$: the temperature limits for a \textit{pure
  thermal wind} with our parameters, shown here with the red,
  dashed-line contours. The excess of viable winds with $T < T_{\rm
  low}$ occurs where cosmic-ray pressure helps drive the wind even at
  low temperatures.  The best fit model for both the ``R4'' and ``R5''
  bands is shown as the gold cross; note that without cosmic-ray
  pressure, no such wind would be possible from the Milky Way.  The
  area where $\chi^2 < 2 \cdot \chi^2_{\rm min}$ is shown as the green
  ellipse.  For winds in this survey, $R_0$, $z_0$, $P_{\rm c,0}$,
  $B_0$, $z_{\rm break}$, and $\alpha$ (see Table~\ref{bestFitParams})
  were fixed.
\label{mDotPlot}}
\end{center}
\end{figure*}

Figure~\ref{mDotPlot} presents the range of mass outflow rates given
in a variety of thermally and cosmic-ray driven wind models.  First,
it is instructive to consider the envelope of winds that successfully
escape the Galaxy's potential vs. the unshaded area where winds could
not be launched.  Towards the bottom-right of the plot, at low
densities and high pressures, gas is hot enough to escape simply by
virtue of $c_{s,0} > v_{\rm esc}$.  These are not outflows that our
code models, and so those regions are not filled-in on the contour
plots.  In fact, if the wind were purely driven by thermal pressure,
gas with $T > T_{\rm high}$ in the plot would ``evaporate'' in this
way.

In the same limit of a thermal-pressure dominated wind, if $T < T_{\rm
low}$, the gas would not have enough energy to escape. Thus, the
upper limit of the shaded area in Figure~\ref{mDotPlot} represents
winds that are becoming too dense (and thus too cold) to escape the
Galactic potential. The ``excess'' of allowed winds with $T < T_{\rm
low}$ exists because of the added cosmic-ray pressure gradient.  The
cosmic-ray pressure component acts over larger distances than the
thermal-pressure component (as $\gamma_c < \gamma_g$), and helps drive
the wind where a thermal wind alone would fail: \textit{cosmic-ray
pressure thus markedly increases the parameter space where viable
winds may be launched}.
%As the comparatively dense and cool winds in the
%unshaded, upper-right section of Figure~\ref{mDotPlot} try to escape
%the potential, they are ``racing'' to escape the potential before the
%thermal pressure drops (see Fig.~\ref{crWindPressurePlot}).  Because
%thermal pressure drops more quickly than cosmic-ray pressure, these
%winds lose their pressure driving before they have escaped the Galaxy,
%and the winds fail: the velocities drop to zero, and the gas will
%begin to fall back towards the Galactic disk.  In contrast, winds with
%a higher fraction of the pressure supplied by cosmic rays can survive
%at large distances (again, near $P_{g,0} \sim 2 \times
%10^4$~cm$^{-3}$~K): the cosmic-ray pressure acts over larger distances
%and in addition, the damping of cosmic-ray generated waves adds
%thermal energy to the gas, and aids in launching the wind to large
%distances.  

%We have tested this explanation by varying both $\gamma_g$ and
%$\gamma_c$; as expected, cool winds that failed for $\gamma_g = 5/3$
%succeed when $\gamma_g = 4/3$.  Meanwhile, for cool winds where cosmic
%rays play a substantial role ($P_{g,0} \sim 2 \times
%10^4$~cm$^{-3}$~K), modifying $\gamma_c = 4/3$ to $5/3$ results in a
%failed wind.  Thus, the apparent difficulty of launching winds at high
%gas pressure is actually the result of not having a correspondingly
%high, slower-decaying cosmic-ray pressure to help drive such high
%$\dot{M}$ winds to large distances.  As expected, cosmic-ray pressure
%thus increases the parameter space where viable winds may be launched.

Looking inside the perimeter of the shaded region in
Figure~\ref{mDotPlot}, we see that, as expected, as we increase the
temperature of the gas (increasing $P_{g,0}$ at constant $n_0$), the
mass outflow rate increases.  Meanwhile, increasing $n_0$ at constant
$P_{g,0}$ decreases the temperature of the gas, yielding less energy
to the gas, and so produces smaller mass outflow rates.

In more detail, though, why does Figure~\ref{mDotPlot} show curvature
in the contour levels?  For any given value of $P_{g,0}$, there are
two values of $n_0$ where a given mass outflow rate can be achieved.
To understand this, recall that the $\dot{M}$ contours are basically
contours of $n_0 v_0$.  The initial velocity, $v_0$, decreases as the
base temperature decreases.  One can think about this as follows: a
decrease in temperature yields a decrease in the energy available to
the gas at the base of the wind, and drives the mass outflow rate
down; so, at some fixed $P_{g,0}$, as the density increases, $v_0$
must decrease.  Since $\dot{M} \propto n_0 v_0$, even as $n_0$
increases, this decrease in $v_0$ leads to a decrease in $\dot{M}$,
and produces the curvature in the $\dot{M}$ contours.

Our chosen (fixed) parameters can also affect the available parameter
space of escaping winds.  For instance, $\alpha \ga 2$ is important
for launching a wind; for $\alpha \sim 1.0$ the wind cannot pass
through a critical point.  Thus, the value of $\alpha$ can also
strongly affect where winds are allowed.  We retain $\alpha = 2$ as in
\citetalias{BrMcVo1991}.

%\begin{figure*}
%\begin{center}
%\includegraphics[width=14cm,angle=90]{figures/windTemperatureMap.ps} 
%% the above is from parameter survey 1, model 2192
%\caption{Temperature at the base of the wind models, in Kelvin.
%\label{tempMap}}
%\end{center}
%\end{figure*}

\subsection{Total Energy Flux in the Wind}\label{totalEnergyFlux}

Of great interest for the production and impact of these winds is
their total power (total rate of energy release in both kinetic energy
and enthalpy).  
%We show the total power (calculated at the base of the
%wind) of a range of cosmic ray and thermally-driven winds in
%Figure~\ref{totalPowerPlot}.  
%At the base of the wind, the enthalpy dominates the kinetic power, so
%the total power simply scales as the base temperature in the wind.  
%At
%the highest pressure levels ($P_{g,0} \sim 10^5$~cm$^{-3}$~K), the
%total power is slightly higher because for those high-pressure winds,
%the added enthalpy of the increased gas pressure increases the total
%enthalpy.  This is not as important at lower pressures due to the
%constant enthalpy in cosmic rays.  
%Approximately $75\%$ (and $\sim 90\%$ for the best-fit model) of the
%power at the base of the wind is in enthalpy.
%Of course, this balance of power would change with height
%in the wind; the calculation here presents the total power at the base
%of the wind, as our chief interest is in whether the Milky Way could
%supply this power.
At the base of the wind, the enthalpy dominates the
kinetic power, so the total power simply scales as the base
temperature in the wind (although plateauing at low gas pressure
because of the added enthalpy in cosmic rays).  Approximately $75\%$
(and $\sim 90\%$ for the best-fit model) of the power at the base of
the wind is in enthalpy.

We now check whether that power can be supplied by supernovae in the
Galaxy.  The total power required for the wind is $\sim 3.7 \times
10^{41}$~ergs~s$^{-1}$.  To calculate the normal SN power in the
Galaxy, we use the SN rates per unit area for both Type I and Type II
SN in \citet{Ferriere01} in her Equations~14 and 16.  Integrating over
the area of the wind only, we find a SN rate of $\sim
5400$~Myr$^{-1}$, which equates to roughly 1 SN every 180~yrs.  If
each SN produces $10^{51}$~$\epsilon$~ergs (where $\epsilon$ is the
fraction of SN power placed in cosmic rays and thermal gas), then the
total SN power in the disk (below the wind only) is $\sim 1.7 \times
10^{41}$~$\epsilon$~ergs~s$^{-1}$.  This simple estimate therefore
shows that the wind requires of order the normal SN rate in the disk,
although it is a factor of $\sim 2$ too high.
%(There is some debate about the SN rate in the Galaxy
%\citep{Ferriere01}, but using the H~\textsc{II}-region derived SN rate
%of \citet{McKeeWilliams97} gives only $\sim 20\%$ higher power.)  
Of course, we are applying a very simple model, and it is quite
conceivable that, by including conduction (which would act as a heat
source to add energy to the base of the wind), and by considering the
effects of clumping within the wind, the outflow's energy requirements
could be reduced (see \S\ref{Conclusions}).
%\clearpage
%\begin{figure*}
%\begin{center}
%\includegraphics[width=10cm,angle=90]{f11.ps} 
%\caption{Total power (kinetic plus enthalpy power) at the base of the
%wind.  As the total energy at the base of the wind is dominated by
%thermal energy, the lines of constant total power largely follow the
%contours of temperature, given by the solid black contour lines.  The
%total power required for the best-fit wind model is within a factor of
%two of the total power supplied by the currently observed SN rate in
%the Milky Way.  For winds in this survey, $R_0$, $z_0$, $P_{\rm c,0}$,
%$B_0$, $z_{\rm break}$, and $\alpha$ (see Table~\ref{bestFitParams})
%were fixed.
%\label{totalPowerPlot}}
%\end{center}
%\end{figure*}
%\clearpage
\subsection{The Importance of Cosmic Rays}\label{cosmicRayDriving}

Figure~\ref{mDotMapCR} plots the mass outflow rate in the wind, as in
Figure~\ref{mDotPlot}, but as a function of cosmic-ray pressure and
gas pressure.  The solid black contour lines in the plot show lines of
constant total pressure.  Thus, moving counter-clockwise along those
lines moves from gas-pressure dominated winds to cosmic-ray dominated
winds.

Before we consider the effect of cosmic rays in these particular
models, we start with some general considerations that will help us
later on.  First, as has been known for some time, momentum addition
either before or after the critical point of a wind affects the
outflow differently \citep[e.g,][]{LeerHolzer80,FSV99}.  Momentum
input before the critical point results in an increase in the mass
outflow rate, whereas momentum input after the critical point
increases the terminal velocity.  In the context of these models, the
smaller adiabatic index of cosmic rays means that the cosmic-ray
pressure decays more slowly (see Fig.~\ref{crWindPressurePlot}), and
therefore increases in the cosmic-ray pressure tend to increase the
terminal velocity as well as the mass outflow rate (although cosmic
rays are less efficient at ``mass-loading'' the wind, or setting the
mass outflow rate, than gas pressure, as we will see shortly).  On the
other hand, increases in the gas pressure tend to increase the mass
outflow rate, with relatively minor changes to the terminal velocity
unless the gas pressure is dominant.  This is an important point to
keep in mind as we discuss the role of cosmic rays in these winds.

Now consider the outline of the shaded region in
Figure~\ref{mDotMapCR}: this outline denotes the transition from wind
models that pass through a critical point and winds that do not have a
critical point, or that fail to launch gas to large heights
altogether.  Again, at the high thermal pressure limit (and therefore
high-temperature limit, near $P_{g,0} \sim 10^5$~cm$^{-3}$~K), the gas
is initially too hot and forms a more evaporative flow.  At high
cosmic-ray pressure towards the top of the map ($P_{c,0} \sim 2 \times
10^5$~cm$^{-3}$~K), we see that cosmic-rays can launch winds even down
to fairly low thermal pressures, if the cosmic-ray pressure dominates
(more on this below).  However, there is an upper limit in cosmic-ray
pressure beyond which the mixed cosmic ray and thermal pressure can
launch an outflow that does not require passage through a critical
point. Beyond this limit, gas of all initial velocities can go to
infinity, and no single wind solution is preferred.

\subsubsection{The Relative Efficiency of Cosmic-Ray Driving}

Looking at the basic structure of colored contour lines in
Figure~\ref{mDotMapCR}, it is clear that generally, as we increase
either the gas pressure or the cosmic-ray pressure, the mass outflow
rate generally increases.  This makes sense since increased pressure
leads to increased energy in the gas which can help increase
$\dot{M}$.

But, it is important to note that the colored contour outlines do not
exactly follow the contours of total pressure.  This is clearly seen
at the high-pressure limit, where the outline of the shaded region
does not follow the contours of total pressure.  This is explained by
the relative inefficiency of cosmic-ray pressure vs. gas pressure in
driving gas from the Galactic midplane.  An intuitive explanation of
this comes from understanding that $(v + v_A) \nabla p_c$ is the rate
that the cosmic-ray pressure transfers energy to the flow ($v \nabla
p_c$ is the rate of work done on the gas moving at $v$, whereas $v_A
\nabla p_c$ is the rate of work done to generate \alfven waves via the
streaming instability).  For $M_A \ll 1$, $v \ll v_A$, and this rate
is much smaller than downstream in the wind where $v > v_A$.  As the
low-velocity region is the region of mass-loading, it follows that
$\dot{M}$ will drop if the proportion of cosmic-ray pressure increases
for some given, constant total pressure.  This explains why following
one contour of constant pressure (counter-clockwise) from high thermal
pressure to high cosmic-ray pressure leads to a drop in mass outflow
rate.

In more mathematical detail, we can see the reason for this difference
in efficiency by considering the drop in cosmic-ray pressure with
density (and therefore drop with height in the wind,
Eq.~\ref{crPressureEq}) in the sub-Alfv\'enic and super-Alfv\'enic
regimes:
\begin{eqnarray}
\lim_{M_A\to0} \frac{dP_c}{dz} & = & \frac{\gamma_c P_c}{2 \rho}
\frac{d\rho}{dz} \\ {\rm and} \lim_{M_A\to\infty} \frac{dP_c}{dz} & =
& \frac{\gamma_c P_c}{\rho} \frac{d\rho}{dz}.
\end{eqnarray}
Since we assume $\gamma_g = 5/3$ and $\gamma_c = 4/3$, the factor of 2
in the the first equation above only exacerbates the difference
between cosmic-ray and thermal driving: $\gamma_c$ is effectively 2/3
compared to $\gamma_g = 5/3$ in sub-Alfv\'enic gas.  The ratio of the
effective $\gamma$ factors is 2.5; this will become important in the
discussion below.  So, in the sub-Alfv\'enic regime, $dP_c/dz$ drops
more slowly than $dP_g/dz$, which means that less momentum is imparted
to the gas; hence, we may conclude that cosmic rays drive gas less
efficiently when the wind is sub-Alfv\'enic.  If the winds were
everywhere (in Fig.~\ref{mDotMapCR}) launched sub-Alfv\'enic, then the
cosmic-ray driven winds would require pressures about a factor of 2.5
higher than the thermally-driven winds.

%This realization is important because it means that even if the
%absolute cosmic-ray pressure is high, compared to the thermal
%pressure, it is still difficult for cosmic rays alone to drive a wind.
%This is why a wind dominated by thermal pressure can be launched with
%$P_{g,0} \sim 4 \times 10^4$~cm$^{-3}$~K, whereas a wind dominated by
%cosmic-ray pressure requires $P_{c,0} \sim 10^5$~cm$^{-3}$~K (see
%Fig.~\ref{mDotMapCR}).  For other values of low thermal pressure and
%lower cosmic-ray pressure (the left side of Fig.~\ref{mDotMapCR}),
%winds cannot be successfully launched from the Galactic midplane for
%the initial gas density considered here (cosmic-ray driven winds would
%be possible for lower initial densities, however).  This difficulty in
%launching is due to two effects.  

This complication is that, at high $P_{c,0} \sim 10^5$~cm$^{-3}$~K,
the winds are somewhat super-Alfv\'enic at the base, whereas at lower
$P_{c,0}$, the winds are Alfv\'enic to sub-Alfv\'enic
(Fig.~\ref{mDotMapCR} is for fixed $n_0$, so the colored contours on
this plot are curves of $v_0$, which therefore decreases as $P_{c,0}$
decreases).  Winds that are launched with super-Alfv\'enic velocities
will have more efficient cosmic-ray driving.  So, some of the
difference in required launching pressures between cosmic-ray and
thermally dominated winds are also due to a transition from
sub-Alfv\'enic to somewhat super-Alfv\'enic which compensates for the
fact that the winds are not strictly launched in the $M_A \ll 1$ or
$M_A \gg 1$ regimes.  

In understanding the relative efficiency of cosmic-ray driving, we
have only explained the basic trends in $\dot{M}$ with total pressure
and the high-pressure limit of Figure~\ref{mDotMapCR}.  We now turn to
the low-pressure limit, where winds also cannot emerge from the
Galactic midplane.

%Interestingly, as can be seen from
%Figure~\ref{mDotMapCR}, the mass outflow rates in these two cases are
%approximately the same, again indicating the relative efficiency of
%cosmic-ray and thermal driving in these winds.

\subsubsection{The Impact of \alfven Wave Damping}

The relative efficiencies of cosmic-ray and thermal pressure driving
do not explain the lack of winds at low thermal pressure ($P_{g,0} <
10^4$~cm$^{-3}$~K) and low to intermediate cosmic-ray pressure
($P{c,0} < 10^5$~cm$^{-3}$~K).  At the very lowest total pressures
($P_{\rm total} \la 10^4$~cm$^{-3}$~K), winds cannot be launched
because the pressure is simply too low (cosmic-ray driven winds would
be possible for lower initial densities, however).  But at higher
cosmic-ray pressures, winds cannot be launched because \alfven wave
damping and the resultant heating dominates.  For intermediate
$P_{c,0}$ and low $P_{g,0}$, the damping of \alfven waves can actually
lead to a large increase in temperature and therefore an
\textit{increase} in $P_g$ with height; this pressure increase can
cause the wind to stall \citepalias{BrMcVo1991}.  So, in this regime,
the \alfven wave damping actually ``pressurizes'' the Galactic disk
and \textit{prevents} a wind from being launched from the midplane.
We have verified this result by running a parameter survey identical
to that shown in Figure~\ref{mDotMapCR}, but with wave dissipation
turned off; as expected, the lack of dissipation allows winds to form
with $P_{g,0} < 10^4$~cm$^{-3}$~K and with $P{c,0} <
10^5$~cm$^{-3}$~K.  For winds with higher $P_{c,0}$ and low $P_{g,0}$,
the cosmic-ray pressure dominates to such an extent that the wave
damping does not hamper wind driving.  (Of course, all of these
constraints on pressure components are only strictly valid for
launching from the Milky Way's midplane, as we have assumed.)

\subsubsection{Launching Winds at Low $P_{c,0}$}

\begin{figure*}
\begin{center}
\includegraphics[width=10cm,angle=90]{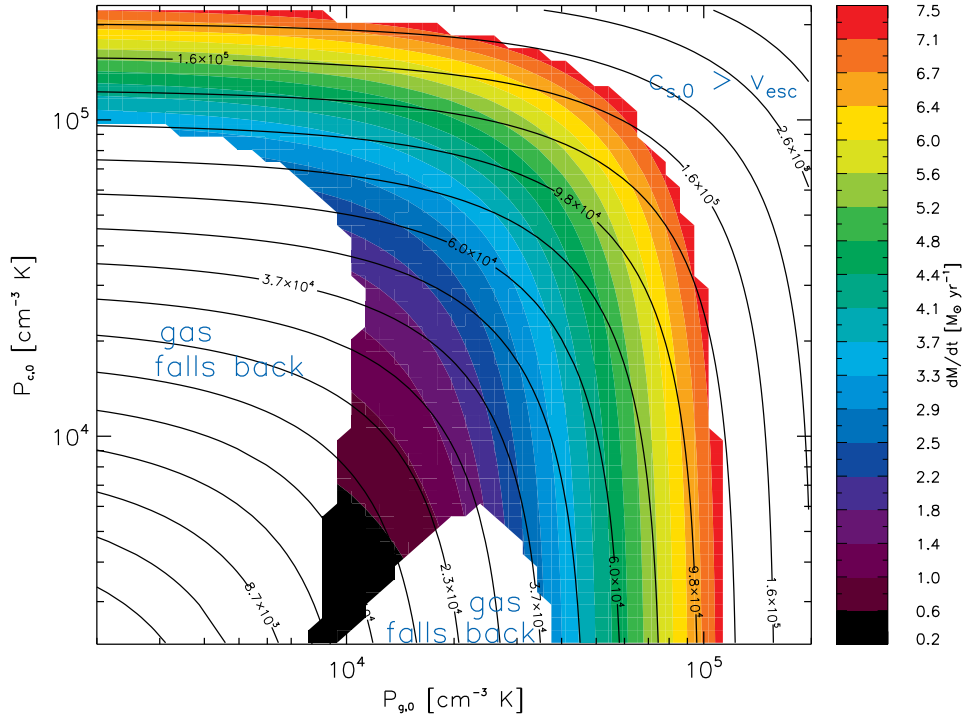}
\caption{Mass loss in the wind vs. changes in the initial cosmic-ray
  pressure and gas pressure.  The black contour lines represent lines
  of constant total pressure.  The winds in the shaded region of the
  contour plot represent mass loss in outflows that pass through a
  critical point; the unshaded regions show locations of parameter
  space where winds cannot be launched (where the total pressure is
  too low) or ``evaporative'' flows that do not pass through a
  critical point (at very high total pressure).  For winds in this
  survey, $R_0$, $z_0$, $n_0$, $B_0$, $z_{\rm break}$, and $\alpha$
  (see Table~\ref{bestFitParams}) were fixed.\label{mDotMapCR}}
\end{center}
\end{figure*}

We now understand most of Figure~\ref{mDotMapCR} except for the small
gap in wind models at low cosmic-ray pressure and intermediate gas
pressure ($10^4$~cm$^{-3}$~K~$\la P_{g,0} \la 4 \times
10^4$~cm$^{-3}$~K, $P_{c,0} \la 5 \times 10^3$~cm$^{-3}$~K).  In this
region, the winds are not so strongly dominated by either cosmic-ray
pressure or gas pressure, \alfven wave damping will not be important
(because of the relatively low $P_c$), and the two must operate in
concert.  The unshaded region defined above is essentially the region
where the inequality in the pressures leads to difficulty in
self-consistently launching a wind (which gas pressure is especially
good at; see above) and driving it to infinity (where cosmic-ray
pressure starts to dominate).  If the thermal wind attempts to load
too much mass at the base of the wind (more than the very low
cosmic-ray pressure can handle), the wind fails at large radii.  If
the thermal wind loads the wind with a fairly low mass outflow rate,
the very low cosmic-ray pressure can continue to loft the gas to large
distances after the gas pressure decays away.  This is corroborated by
the very low mass outflow rate associated with the ``promontory'' of
low mass outflow rate near $P_{g,0} = 9 \times 10^3$~cm$^{-3}$~K and
$P_{c,0} = 2 \times 10^3$~cm$^{-3}$~K.

%The increase in mass outflow rate is not quite that simple, though; it
%does not follow the contours of constant total pressure, as one might
%expect.  The relative inefficiency of cosmic-ray pressure in driving
%gas from the Galactic midplane, as discussed above, can explain this.
%Following one contour of constant pressure (counter-clockwise) from
%high thermal pressure to high cosmic-ray pressure leads to a drop in
%mass outflow rate, due to the lower efficiency of cosmic-ray driving.

%Cosmic rays are also very important in determining the terminal
%velocity in our wind models, $v_{\infty}$.  As mentioned previously,
%the slow-decay of cosmic-ray pressure (due to our assumption that
%$\gamma_c = 4/3$) leads to cosmic-ray pressure dominating beyond the
%critical point, which helps define the terminal velocity in the wind.
%Therefore, $v_{\infty}$ increases as $P_{c,0}$ for values of $P_{c,0}
%\ga P_{g,0}$.  For cases where the gas pressure dominates, $P_{g,0}$
%also helps set $v_{\infty}$.

%\begin{figure*}
%\begin{center}
%\includegraphics[width=14cm,angle=90]{figures/windVInfCR.ps} 
%\caption{The terminal velocity of the wind (here, recorded at $z =
%  1$~Mpc) as a function of cosmic-ray pressure and gas
%  pressure.\label{vInfMapCR}}
%\end{center}
%\end{figure*}

%\begin{figure*}
%\begin{center}
%\includegraphics[width=14cm,angle=90]{figures/windTotalPowerCR.ps} 
%\caption{The total wind power (kinetic power plus enthalpy in the
%  wind) as a function of cosmic-ray pressure and thermal
%  pressure.\label{totalPowerMapCR}}
%\end{center}
%\end{figure*}

\section{Conclusions}\label{Conclusions}

We have implemented a simplified cosmic ray- and thermally-driven wind
and have used it to try to explain the soft, diffuse X-ray emission
seen towards the Galactic Center.  We find that such a wind can indeed
match the observed averaged X-ray emission quite well, and in fact
fits demonstrably better than the static polytrope model of
\citet{AlmyEtAl00}.  It is important to note that this wind is
approximately equally powered by both cosmic rays and thermal
pressure: cosmic rays are important in helping this relatively cool
wind escape from the Galactic potential.  It is also quite interesting
that this wind does not require excessive thermal or cosmic-ray
pressures (both pressures are not extreme compared to what has been
estimated for the inner Milky Way), nor does this simple model require
much more energy than the standard inferred supernova rate implies.
Taking this result at face value, such a wind would be very important
to the ``ecology'' of the Milky Way due to the high mass loss rate of
2~$M_{\odot}$~yr$^{-1}$.  In addition, such a wind would also play an
important role in removing angular momentum from matter in the
Galactic disk and allowing matter to move radially inward
\citep{Zirakashvili1996}.  At the least, this shows that such wind
models should be considered further for the Milky Way; they may be
able to explain at least a substantial fraction of the observed soft
X-ray emission.

Further, as other researchers have already shown \citep{BrDoVo02},
such winds can also be used to explain the unexpectedly slow rise in
$\gamma$-ray emission towards the center of the Galaxy; this work
therefore gives independent support to the Galactic wind hypothesized
in \citet{BrDoVo02}.  We have not yet calculated the effect of the
best-fit wind model on the cosmic-ray distribution, but a simple
approximation will investigate if this wind is removing cosmic rays at
too high a rate.  We reason as follows.  The best-fit wind model shown
here has a cosmic-ray advective timescale of $\sim 4.3 \times
10^6$~years.  Therefore, supernova in the disk must resupply the
cosmic-ray pressure on that timescale.  We know that the approximate
total supernova energy in the disk (from observations, see
\S\ref{totalEnergyFlux}) is $\sim 1.7 \times 10^{41}$~ergs~s$^{-1}$.
So, if some fraction of this energy, $\epsilon_{\rm CR}$, is given to
cosmic rays, and distributed over the volume occupied by hot gas where
the wind is launched (over the Galactocentric radius range of $1.5$ to
$4.5$~kpc, and a height range of $\pm$2~kpc, to be conservative), that
energy density should be similar to the cosmic-ray pressure required
to launch the wind.  Calculating the resultant buildup of $P_{\rm c}$
over $4.3 \times 10^6$ years at the observed SN rate, we find $P_{\rm
c} \sim \epsilon_{\rm CR} \cdot 6.9 \times 10^{-12}$~dyne~cm$^{-2}$ or
$\sim \epsilon_{\rm CR} \cdot 5.0 \times 10^4$~K~cm$^{-3}$.  This is
actually of order the $P_{\rm c}$ that the best-fit model requires
(see Table~\ref{bestFitParams}), although it would require
$\epsilon_{CR} \sim 0.6$ to duplicate the best-fit $P_{\rm c}$, which
is relatively high.  Still, this simple, rather conservative
calculation shows that the high $\dot{M}$ wind shown here does not
remove cosmic rays much more quickly than they can be replenished by
the normal SN rate in the Galaxy, although the removal rate of cosmic
rays is certainly non-negligible, and would affect the density of
cosmic rays towards the Galactic center.  Of course, a more detailed
calculation is required (with a more detailed wind model), but this
again shows that the best-fit wind model is at least feasible, and
would have a significant but not destructive effect on the Galaxy's
cosmic-ray density.

\subsection{Future Improvements}

More detailed models are clearly needed; there are a few concerns
about the current model that could be addressed with more realistic
wind models.  For instance, we have assumed uniform density at the
base of the wind over the area of the disk from $R=1.5$~kpc to
4.5~kpc, which leads to a mass outflow rate of order
2~$M_{\odot}$~yr$^{-1}$.  This seems quite high, but in the context of
a more detailed model with variations in density within the wind, we
might expect that the $n^2$ weighting of emissivity would favor
overdensities, and allow an inhomogeneous wind to better reproduce the
observations with a smaller mass outflow rate.  In addition, it is
possible that other effects limit the gas to velocities below those in
this simple model; drag effects may slow down the wind
\citep[e.g.,][]{EM07}, and could lead to some of the gas forming part
of the Galactic fountain \citep{Bregman99}.  On the other hand,
turbulence may be an important additional source of energy for the
wind, but we have not included such an input in this work.  Also, the
effects of distributed mass loading, which could be relatively easily
incorporated into this model, have been ignored so far; such mass
loading would be very important to the emission properties of the
wind, and the potential observability of such winds in many galaxies.
Finally, note that we have used \citet{AG89} abundances; if the wind
starts out with super-solar abundances (particularly in oxygen), a
smaller mass outflow rate would be required.

The cosmic ray physics in this wind is still quite simple.  As
mentioned previously, we have assumed zero diffusivity of the cosmic
rays throughout the wind \citep[c.f.,][]{BrMcVo1993}.  We have also
neglected thermal conductivity (our calculations show that
conductivity may be important for $z \la 350$~pc).  Including both of
these effects would be essential to future progress for this wind
model: as mentioned previously (see \S\ref{totalEnergyFlux}),
including the effects of conduction may help lower the energy
requirements in the wind, bringing the total power of the wind closer
to that of the total inferred supernovae power in the Galactic disk
below the wind.  We also note that we are using a model of the
gravitational potential that we know overestimates the potential in
the Milky Way (see \S\ref{modelDef}).  Adopting the more detailed
potential model of \citet{DehnenBinney98} may allow a lower
total-energy wind to duplicate the soft X-ray observations.  Finally,
we have also assumed the hydrodynamic model of \citet{McKenzieWebb84}
and \citet{BrMcVo1991} for the interaction of cosmic rays with \alfven
waves and the gas.  To further examine this model, we will next
consider the effects of higher cosmic-ray fluxes \citep{Zweibel2003}
and apply it in other settings.

\subsection{Future Tests}

How can we further test this model?  In analogy to early studies of
the solar wind, this outflow may impact clouds in the vicinity of the
galaxy, perhaps causing ``comet-tail'' extensions to high velocity
clouds above the plane of the Milky Way.  The formation of such
``tails'' would depend on the velocity of the wind.  This has been
studied in some detail before \citep{BenjaminCox02}, but should be
reconsidered in the context of the predictions of these winds.  In
addition, it may be possible to study the kinematic impact of this
wind on, for instance, the Magellanic Stream (A. Burkert, personal
communication).

Another way to test the model would be to compare the wind with
absorption columns and emission spectra towards the center of the
Galaxy.  Concerning absorption measurements, recent \textit{Chandra}
observations \citep{FutamotoEtAl04} towards the low-mass X-ray binary
4U 1820-303 show significant columns in \al{O VII}, \al{O VIII}, and
\al{Ne IX}.  As this X-ray binary is located within ten degrees of the
Galactic center, at a distance of approximately 7.6~kpc, it seems an
ideal target.  However, similar absorption columns are found towards
objects on sightlines that do not intercept the base of the wind; for
instance, Mrk~421 shows similar absorption columns
\citep{FutamotoEtAl04}, but is far out of the plane.  This leads us to
conclude that the observed absorption is somewhat local to the Sun's
position, and, as such, this absorption does not constrain the wind
model.  However, recent Suzaku measurements of emission at various
latitudes along different sightlines towards the Galactic center
(Rocks, 2008, in preparation) may help constrain the properties of the
wind.

Finally, these winds, including their important cosmic-ray component,
will also emit synchrotron radiation.  We are now calculating the
synchrotron emission expected from these wind models (Schiller et al.,
2008, in preparation).  This will allow exploration of the wind's
synchrotron emission as compared to recent models of Galactic
synchrotron which begin to map the three-dimensional cosmic-ray
emissivity \citep{NordEtAl06}.

Looking at a wider field of application, such wind models may also be
quite important in application to starburst galaxies
\citep[e.g.,][]{GallagherSmith05, SDR06} and dwarf galaxies.  The
general applicability of these kinds of models to starbursting
galaxies has been shown by \citet{Breitschwerdt03} in fitting
cosmic-ray and thermally driven wind models to NGC 3079.

\acknowledgements
The initial impetus, in our group, for the investigation of a
large-scale Galactic wind came from Dr. Don Cox; we are indebted to
him for bringing the idea to our attention, and for various helpful
conversations as the model was being developed.  We thank the referee,
Dr. Dieter Breitschwerdt, for his thorough reading of the paper and
insightful comments and questions.  We also thank Dr. Richard Almy for
the development of the initial version of the code to map X-ray
emission from the wind for comparison with \textit{ROSAT} data.
Finally, we thank Sebastian Heinz and Andreas Burkert for helpful
comments and conversations.  This work was supported by NSF
AST-0507367 and NSF PHY-0215581 (to the Center for Magnetic
Self-Organization in Laboratory and Astrophysical Plasmas) and NASA
ATP grant NAG5-12128 (RAB).  This research has made use of NASA's
Astrophysics Data System.

\newcommand{\noopsort}[1]{}

\end{document}